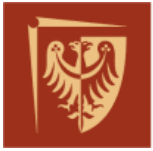 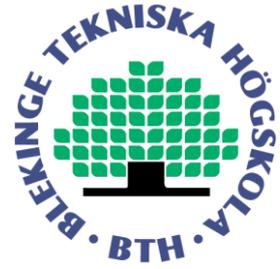

# Key User Extraction Based on Telecommunication Data

**Piotr Bródka**

Institute of Informatics, Faculty of Computer Science and Management, Wrocław University of Technology, Wyb. Wyspiańskiego, 50-370 Wrocław, Poland

School of Engineering Blekinge Institute of Technology Box 520 SE – 372 25 Ronneby Sweden

**Wrocław/Ronneby 2008**

This thesis is submitted to the Wrocław University of Technology and Blekinge Institute of Technology within Double Diploma program for the degree of Master of Science in Computer Science.

**Contact Information:**
Author(s):
Piotr Bródka
E-mail:     piotr.brodka@gmail.com
            piotr.brodka@pwr.wroc.pl

Wrocław University of Technology advisor          Blekinge Institute of Technology advisor:
**Przemysław Kazienko**                           **Ludwik Kuzniarz**
Institute of Informatics                          School of Software Engineering

# ABSTRACT

The number of systems that collect vast amount of data about users rapidly grow during last few years. Many of these systems contain data not only about people characteristics but also about their relationships with other system users. From this kind of data it is possible to extract a social network that reflects the connections between system's users. Moreover, the analysis of such social network enables to investigate different characteristics of its members and their linkages. One of the types of examining such network is key users extraction. Key users are these who have the biggest impact on other network members as well as have big influence on network evolution. The obtained about these users knowledge enables to investigate and predict changes within the network. So this knowledge is very important for the people or companies who make a profit from the network like telecommunication company. The second important thing is the ability to extract these users as quick as possible, i.e. developed the algorithm that will be time-effective in large social networks where number of nodes and edges equal few millions.

In this master thesis the method of key user extraction, which is called social position, was analyzed. Moreover, social position measure was compared with other methods, which are used to assess the centrality of a node. Furthermore, three algorithms used to social position calculation was introduced along with results of comparison between their processing time and others centrality methods.

## *Streszczenie (Polish)*

Liczba systemów gromadzących dane o swoich użytkownikach gwałtownie wzrosła w czasie ostatnich kilku lat. Wiele z tych systemów zawiera dane o ogromnej liczbie użytkowników. Na podstawie tych danych można stworzyć sieć społeczną i analizować ją. Jednym z typów analiz jest wydzielanie użytkowników kluczowych dla sieci, którzy mają na nią największy wpływ i biorą ogromny udział w jej formowaniu ewolucji. Mając informację, kim są ci użytkownicy możliwa jest ich obserwować i na tej podstawie przewidywać zmiany w sieci lub też wywierać wpływ na nich by takie zmiany wywołać. Więc wiedza ta jest bardzo ważne w przypadku ludzi lub firm, które czerpią zyski z sieci np. firmy telekomunikacyjne. Inną ważną rzeczą jest by wydzielanie tych użytkowników wykonać w jak najkrótszym czasie.

W pracy zaproponowano sposób wydzielania użytkowników kluczowych przy pomocy metody zwanej Pozycją Społeczną. Następnie zaprezentowano wyniki jej użycia oraz porównano ją z innymi metodami. Pokazano też trzy metody obliczania pozycji społecznej, pokazano, która z nich jest najszybsza oraz porównano ich szybkość z najprostszymi metodami analizy sieci społecznych.

# CONTENT

# 1 INTRODUCTION

A social network is a social structure that consists of nodes where node is a single social entity i.e. a person, a group of people, an organization. Nodes can be tied by different kinds of relations, like financial exchange, friendship, hate, love, trade, web links, or airline routes [25]. The representation of a social network can be e.g. a graph in which nodes represent people, organizations or other social entities that are connected with ties – edges of the graph.

Social network analysis is used to analyse social networks [46], [49]. Social network analysis has been created based on the graphs theory as the mathematical instrument for social networks' interpretation. Traditional social analysis focuses on users, i.e. their attributes, features, etc., while social network analysis focuses on connections. However, user can be analysed, on a second, third or even last step of whole process. To analyse the connections between users many different measures were created and are still being developed (e.g. centrality, prestige, social position, social capital, etc.) [19], [27].

During past few years the number of social services available in the Internet such as Flickr, YouTube, Facebook or Friendster has grown rapidly. Those services are the basis to extract the social networks where people can share their files, photos, thoughts, find colleagues, friendships or even love. These kind of social networks are called virtual social networks, online communities or web-based social networks [34], [50].

A social network can be extracted from data available in many multi-users systems where people communicate or cooperate with each another. Such example of a social network can be derived from the telecommunication system [31] as this presented in this thesis.

Based on the telecommunication data (data about phone calls delivered by Big European Telecomunication Provider containing caller id, receiver id, time of phone call, date of phone call and duration of phone call) we are able to create a social network, in which nodes represent particular phone numbers and edges represent relations between these phone numbers. The phone numbers can represent a single user, while edges can be extracted based on the phone calls that were made form one phone number to another.. In such a network may exist important users, who have significant role for the whole network or at least for a part of it. The knowledge about such users allows the telecommunication company or advertisement company to take care of such them, propose some promotion and keep this client in company.

This master thesis shows method of extracting key users from the community on the example of social network that is created based on telecommunication data. The method utilizes the social position measure. Different approaches to the social position calculation and their comparison are shown.

## 1.1 Aim and Objectives

The aim of this master thesis was to gather knowledge about a social network extracted from telecommunication data and find the way to extract key user from this network. Additional objectives were:

1. to gather knowledge about the subject (social network, social position measure) through research in the literature,

2. to conduct research with telecommunication data in order to receive information about network and network users,

3. to find correct subset of data which meets a social network theory and can be transform into the social network,

4. to find the different commitment functions for the social network,
5. to prepare several different approaches to calculate a social position,
6. to compare these approaches with each other and with existing centrality measures,
7. to compare that social position with other measures may be used to extract key users.

## 1.2 Research Questions

The thesis addressing following research questions:

1. Which data from the telecommunication data has to be collected in order to create the relevant from the point of the key users extraction process social network?
2. What form of the commitment function is possible to create using the extracted telecommunication data?
3. What is the difference between various commitment function?
4. How the social position can be calculated?
5. Can the social position measure be utilized to extract key users?
6. What is the difference between various method of social position calculation?
7. What is the difference between the social position measure and other centrality measures?

## 1.3 Expected Outcomes

1. State-of-art of social networks
2. A set of guidelines which shows how to extract key users from telecommunication data
3. A set of tools developed during research which will be helpful during data investigation and key user extraction
4. A thesis containing the knowledge gathered during research and answers for research questions.
5. New algorithms to a social position calculation.
6. Results of the experiments with results discussion.

## 1.4 Research Methodology

In this project both qualitative and quantitative methods was used. Qualitative methods was used during literature study to increase the understanding of the most important concepts like users, connections between users, social networks, social network analysis and user's role in social network. Literature survey allows to produce a good background for further study.

During telecommunication data investigation both qualitative and quantitative methods was used to prepare, clean and analyse it.

Quantitative methods was useful while prototypes building process and during comparison of prototypes Experiments and results discussion also used quantitative methods to investigate proposed solution against thesis objectives.

## 1.5 Chapters Content

The rest of the thesis is organized as follows. Chapter 2 contains introduction and basis of the social networks theory, notation and representation of the social network, description of the social network analysis and measures in the social network analysis, and presentation of the virtual social network, i.e., the whole background for the thesis. Chapter 3 contains theoretical and mathematical basis of commitment function and social position method also three different approaches to social position calculation and three algorithms are described in this chapter. Chapter 4 is a general description of all common SNAP's modules and detailed description of modules which was made by this master thesis author. Chapter 5 contains description of the telecommunication data and the presentation of the process which allow to prepare social network witch two different commitment functions. Chapter 6 describes investigation of the features of social position measure, such as average, minimum and maximum value of SP, distribution of its values, etc. In his chapter the influence of $\varepsilon$ coefficient on the social position value and its characteristics i.e. the mean value of SP, the distribution of SP, minimum and maximum values of SP for each $\varepsilon$, the ranking itself is also presented. Moreover, the comparison of social position measure with indegree and outdegree centrality is described together with the efficiency tests which compared processing time of different variants of SPIN algorithm as well as different centrality indices. Final concludes from experiments carried out and answers for research questions are in chapter 7.

# 2 SOCIAL NETWORK

## 2.1 General Concept of Social Network

First time the term "social network" was used by Barnes in 1954 [4]. According to his definition a social network is a group of people drawn together by family, work or hobby where the size of the group is about 100-150 people.

Nowadays, the definition is more specific "*A social network is a social structure made of nodes (which are generally individuals or organizations) that are tied by one or more specific types of interdependency, such as values, visions, idea, financial exchange, friends, kinship, dislike, conflict, trade, web links, sexual relations, disease transmission (epidemiology), or airline routes.*" [25]. Moreover, many organizations create their own definitions. "*The personal or professional set of relationships between individuals. Social networks represent both a collection of ties between people and the strength of those ties.*" [28] is the definition used by Scrutiny of Acts and Regulations Committee in Australia. Another example: "*A web of interconnected people who directly or indirectly interact with or influence the student and family. May include but is not limited to family, teachers and other school staff, friends, neighbors, community contacts, and professional support.*" [26] used by Rehabilitation Research & Training Center on Positive Behavioral Support – funded by U.S Education Department. This shows that although the concept of social network appears to be quite obvious, almost every organization describes the social network in a slightly different way and in consequence many different variations of social network definition exist. Some of researchers define social network in a very formal way, e.g. Yang, Dia, Cheng, and Lin [54], [42] who claim that social network is an undirected, unweighted graph while the others prefer more sociological approach [49], [22]. Wasserman and Faust define the social network as the finite set or sets of actors and one or more relations defined on them [49] whereas Hatala claims that it is a set of actors with some patterns of interaction or "ties" between them, represented by graphs or diagrams illustrating the dynamics of the various connections and relationships within the group [22]. Garton, Haythorntwaite, and Wellman [16] propose the following definition of social network – it is a set of social entities connected by a set of social relationships. Yet another definition is that presented by Liben-Nowell and Kleinberg, i.e. a scial network is a structure whose nodes represent entities embedded in the social context, and whose edges represent interaction, collaboration, or influence between entities [38].

In this thesis, the following definition was used:

*Social network is a tuple ($M,R$) over a set of actors ($M$). The elements of this relation are called connections ($R$). Actor (member) is a single social entity i.e. a person, a group of people, an organization, a company, a city, a country, etc. Connection (relation) is a relationships, activity or interdependences between two actors.*

Several examples of social networks can be enumerated: a family [6], a friendship network of students [3], a community of scientists or other professionals in the given discipline who collaborate [42] or prepare common scientific papers, a corporate partnership network [37], a set of business leaders who cooperate with one another [38], a company director network [43], a group of acquaintances who share similar interests, etc.

## 2.2 The Small World Phenomenon

In 1960s social psychologist Milgram carried out the first big experiment concerning social networks. As a result, he created *the small world problem* (phenomena) [39]. He picked up one target person who lived in Boston and three groups of starting persons. Each of the starting persons received a letter which included the description of the study, basic information about the target person and the request to send the letter to the receiver through its colleague. Results revealed that average path length from the starting person to the target was 5.2. This study showed there were only 5 or 6 people needed to connect two random people in such a big country as the USA. The second interesting thing is that the target person received 16 mails from his neighbour, 10 from one work colleague and 5 from the second work colleague [12].

This experiment showed that

- Natural social networks has not boundaries

- There exist people in the network who are more important (has higher position) than others.

- Information, gossip, viruses, etc. can spread through the network very quick.

In 2003, Watts, Muhamad, and Dodds carried out the similar experiment but worldwide. They used more than 60,000 e-mail users and 18 target persons in 13 countries. They "*estimate that social searches can reach their targets in a median of five to seven steps*" [13]. It confirms that Milgram's results was correct.

## 2.3 Notation and Representation of Social Network

There are three comprehensive approaches to the representation of social network. Three main types of notations can be distinguished: graph, sociometric, and algebraic approach.

A social network can be represented by one of the mathematical tools that are graphs. The graph theory has been widely studied by many researches [10], [5], [20] and the social network analysis has adopted this method of representation because it is very useful for calculation the centrality and prestige within network, identification of cohesive subgroups, etc. [46], [49]. Flament in 1963 [15] and Harary in 1965 [21] were one of the first scientists who analyzed the usage of graphs in social networks. The basic definition of a graph, and in consequence also a social network $SN=(M,R)$ is as follows: it is a finite set of nodes (network members) $M$ and the set of arcs (relationships) $R$ that connects them [12] (see Figure 1). Such graph $SN$ depending on the character of the connections can be either undirected or directed. The former consists of nodes and arcs that fulfill the condition: for each arc $(m_i,m_j) \in R$: $(m_i,m_j)=(m_j,m_i)$. In other words, in the case of undirected graph, if there is a connection from $m_i$ to $m_j$ then simultaneously exists arc from $m_j$ to $m_i$ [49]. In the directed graph $(m_i,m_j) \neq (m_j,m_i)$. It means that the existence of the connection from $m_i$ to $m_j$ does not entail the existence of the relation $(m_j,m_i)$. [49, 12]. Graphs can be also weighted (also called valued) as well as unweighted. In social network analysis the relations within the unweighted graph are called binary ones, and they indicate only the fact of the existence of the symmetric relation between two nodes. In the weighted graph, its weights denote the strength of the connections (relations) between two nodes (members).

For better understanding of this thesis a few terms need to be introduced:

- **Walk** – a sequence of actors and connections which starts and ends with an actor. Closed walk is a walk which starts and ends with the same actor [45].

- **Trail** – is a walk between two actors which contains the given connection only once (however one actor can be a part of a trail many times). **Length** of the trail is a number of connections it contains [45].

- **Path** – is a walk in which the single actor and single connection can be used only once. The exception is a closed path which starts and ends with the same actor. **Length** of the path is the number of connections it contains. Two paths are independent if their actors sets are disjunctive (they share no actors), only start and end point can be the same [45].

- **Neighbourhood of actor A** – it is a set of all actors which are directly connected with actor A (path length between them and actor A is 1).

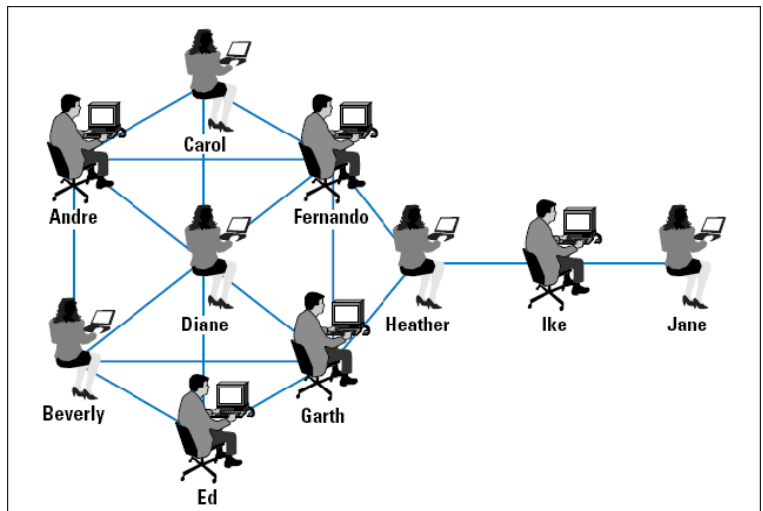

Figure 1 Example of a simple social network. Nodes are people and the edges represent data exchange or information flow [36].

In the sociometric notation a social network is represented by sociomatrix, which is adjacency matrix for graph [49,12]. Sociometric notation, introduced by Moreno [41], is used to study the structural equivalence and blockmodels [49]. In sociomatrix, each line and column corresponds to a node from graph *SN*. The nodes are taken in the same order for both lines and columns. An element of the matrix denotes the fact of the existence of the connection between two nodes and it contains the strength of the relation in case of valued networks. For example, the unweighted and directed graph can be represented by the matrix which elements can have two values: 1 when there is connection from $m_i$ to $m_j$ and 0 when such relation does not exist. The matrix can be either symmetrical when it represents an undirected graph or asymmetrical when it describes the directed graph. Moreover, it will contain only 1 and 0 values when the social network is unweighted one. The sociometric notation facilitates algebraic computations and transformations on matrixes.

Algebraic approach is most appropriate for role and positional analyses, relational algebras, and is used to study multiple relations [49]. This notation is designed for one-mode networks and was first utilized in [53] and [8].

## 2.4 Social Network Analysis

Social network analysis stems from traditional social analysis used by sociologists and anthropologists in the first half of the 20th century. After introducing mathematical interpretation of social networks scientists started developing social network analysis[1].

One of the most popular definition was proposed by Valdis Krebs: *"Social network analysis [SNA] is the mapping and measuring of relationships and flows between people, groups, organizations, computers, web sites, and other information/knowledge processing entities. The nodes in the network are the people and groups while the links show relationships or flows between the nodes. SNA provides both a visual and a mathematical analysis of human relationships."* [36].

The regular social data (Table 1) is quite different than social network data (Table 2). Traditional social data describes actors whereas social network data can contain social data but mainly describes connections between actors rather than actors itself [19].

---

[1] History of social network analysis available at [6].

| Name    | Gender | Age | Marital status |
|---------|--------|-----|----------------|
| **Carol**   | Female | 32  | Married        |
| **Jane**    | Female | 26  | Single         |
| **Richard** | Male   | 30  | Single         |
| **Andre**   | Male   | 45  | Married        |

Table 1 Example of simple social data

| Who likes whom? | | | | |
|---|---|---|---|---|
| **Name A\B** | Carol | Jane | Richard | Andre |
| **Carol**   | - | 0 | 1 | 0 |
| **Jane**    | 1 | - | 0 | 1 |
| **Richard** | 1 | 1 | - | 0 |
| **Andre**   | 1 | 0 | 1 | - |

Table 2 Example of social network data. 0 – person A does not like person B, 1 – person A like person B.

Because of the fact that social network analysis focuses on investigation of connections it does not mean that social network analysis is not interested in actors. After receiving conclusions social analysis may study actors to retrieve additional information and to better understand this network.

In social network analysis four main steps can be distinguished [16], i.e.: selecting a sample, collecting data, choosing and applying the method of social network analysis, drawing conclusions.

In order to identify and investigate the patterns that occur within the network, first the selection of a group of people should be done. The possibility of analyzing every node of the network (especially these huge and heterogeneous) is usually limited by the available resources and because of that the representative group of actors ought to be chosen for further analysis. This group of actors is called population [19] or sample [16]. After that, the data is collected. Many methods of gathering data such as questionnaires, interviews, observation, and artefacts exist [16]. However, most of researches agree that the best method is the hybrid one that copes with the shortcomings of the enumerated methods and combines all of them [44]. The researches distinguish the types of data that should be investigated. The data to analyse also called units of analysis are as follow: relations, ties [16], and actors.

The next step in social network analysis is to choose the most suiting method of analysis. Social network analysis has three approaches to the analysing process (Figure 2):

- **Full network methods** – those methods collect and investigate data about the entire network (each actor and each connection). This approach gives the best results but is the most expensive, very time-consuming and sometimes it is impossible to collect the full data. However, full network methods are necessary to calculate some measures (e.g. betweenness – see section 2.5) [19].

- **Snowball methods** – methods start with one local actor or small set of actors. Each actor have to show some or all his connection to other actors. Actors picked up in second step have to do the same thing like first actors. The whole process ends when no new connections are shown or after the predefined number of iteration. This method is very useful in finding strong connected group in big networks but it has few weakness. Firstly, if a person is isolated or very loosely connected, he or she might be never found by this method. Secondly, if the first actor will not be chosen properly, the method can result with nothing. Because of that the snowball method usually is used after pre-study which locates the good starting point (e.g. president/governor for country or CEO of company) [19].

- **Ego-centric method** – this method investigates only one actor (ego) and his neighbourhood (also connections between his neighbours). This method can provide quite good information about the local network and how this network affects this actor. Additionally, if ego was chosen randomly it gives the incomplete view of the whole network. However, the method is efficient both in time and resources [19].

The last step that enables to identify the existing within the particular social network patterns is to draw the conclusion from the investigation. The issue that has to be emphasized is that collecting network data and picking the right method of analysis is an extremely challenging task.

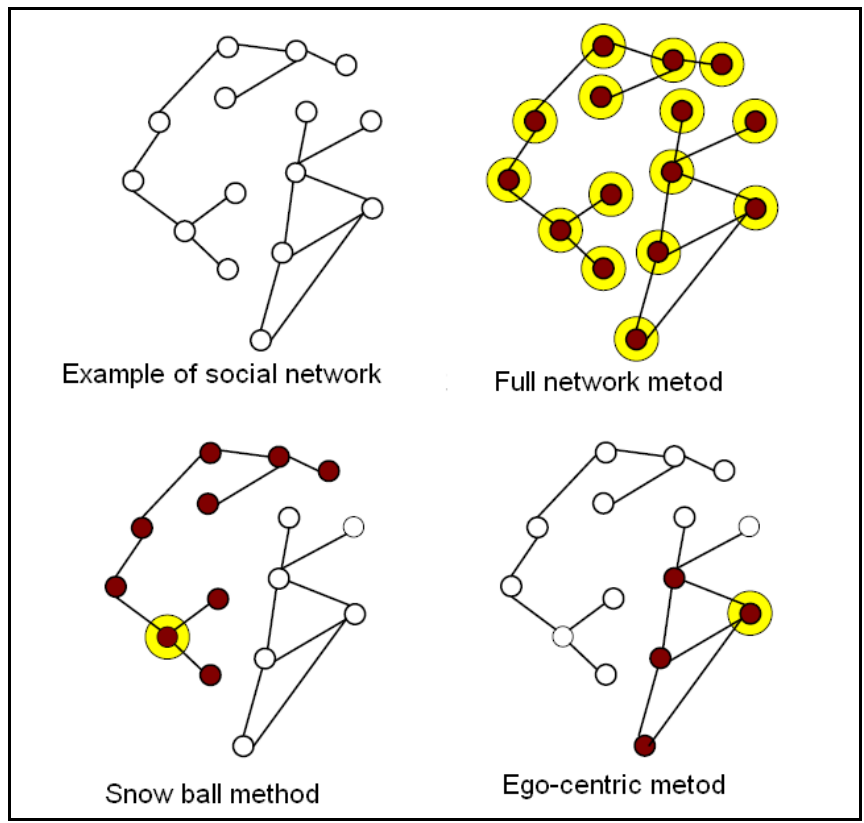

Figure 2 Visualisation of social network analysis methods [33].

Nevertheless, due to its potential, the social network analysis is becoming the main technique in modern sociology, anthropology, sociolinguistics, geography, economics, social psychology, communication studies, information science, organizational studies, and biology [25]. The list of possible applications which use social network analysis is included in Appendix 1.

## 2.5 Measures in Social Network Analysis

Measures (also called metrics) are used in social network analysis to describe the actors' or ties' features, characteristic within social network as well as to indicate personal importance of individuals in social network so this measures can be used to extract key users from the network. In further sections, there is a list of the most popular and useful measures that are utilized to identify the most powerful, important node or a group of nodes in the social network.

## 2.5.1 Centrality Degree

A centrality degree is the simplest and the most intuitive measure among all. It is the number of links that directly connect one node with others. In an undirected graph it is the number of edges which are connected with the single node. In a directed graph, degree is divided in indegree for edges which are directed to the given node and outdegree for edges which are directed from the given node. On the example from Figure 1, Diane has the biggest centrality degree because she has 6 direct ties. A centrality degree is determined using:

$$C_D(x) = d(x) \qquad (1)$$

where $d(x)$ is the number of nodes which are directly connected to node $x$. A centrality degree is normalized using:

$$C_D(x) = \frac{d(x)}{n-1} \qquad (2)$$

where $n$ is the number of nodes in a network. [9, 12, 46, 23].

Table 3 presents the centrality degree ($C_D$) values for the social network from Figure 1.

## 2.5.2 Centrality Closeness

A centrality closeness describes how close a node is to all other nodes in a network and tells how quick this node can reach all other nodes (for e.g. to spread some information to entire network). This measure emphasizes quality (position in a network) rather than quantity (number of links, like in a centrality degree measure). On the example from Figure 1 Fernando and Garth have the best closeness despite having fewer direct ties than Diane. They have "shortest path" and they are closer to others than anyone else. Centrality closeness is determined using

$$C_C(x) = \frac{1}{\sum_{\substack{y \neq x \\ y \in A}} c(x,y)} \qquad (3)$$

where $c(x,y)$ is a function describing the distance between nodes $x$ and $y$ (i.e. max, min, mean or median). Closeness is normalized using

$$C_C(x) = \frac{n-1}{\sum_{\substack{y \neq x \\ y \in A}} c(x,y)} \qquad (4)$$

where $n$ is the number of nodes in a network [9, 12, 23, 36].

Table 3 presents the centrality closeness ($C_C$) values for social network from Figure 1.

## 2.5.3 Centrality Betweenness

A centrality betweenness describes how often node is between two other nodes and how many paths go through this node. Actors with high centrality betweenness are very important to the network because others actors can connect with each other only through them. On Figure 1 without Heather Ike and Jane would be outside of the network. Betweenness for node *n* is counted by

$$C_B(x) = \sum_{\substack{i \neq x \neq j \\ i,j \in A}} \frac{b_{ij}(x)}{b_{ij}} \qquad (5)$$

where $b_{ij}(x)$ is number of shortest paths from *i* to j that pass through *n*, and $b_{ij}$ is number of paths from *i* to *j*. Centrality betweenness is normalized using

$$C_B(x) = \frac{\sum_{\substack{i \neq x \neq j \\ i,j \in A}} \frac{b_{ij}(x)}{b_{ij}}}{n-1} \qquad (6)$$

where *n* is the number of nodes in a network [9, 12, 46, 23, 36].

Table 3 presents the centrality betweenness ($C_B$) values for social network from Figure 1.

| Name\Measure | $C_D$ | $C_C$ | $C_B$ |
|---|---|---|---|
| **Diane** | 0.666 | 0.600 | 0.102 |
| **Fernando** | 0.556 | 0.643 | 0.231 |
| **Garth** | 0.556 | 0.643 | 0.231 |
| **Andre** | 0.444 | 0.529 | 0.023 |
| **Beverly** | 0.444 | 0.529 | 0.023 |
| **Carol** | 0.333 | 0.500 | 0.000 |
| **Ed** | 0.333 | 0.500 | 0.000 |
| **Heather** | 0.333 | 0.600 | 0.389 |
| **Ike** | 0.222 | 0.429 | 0.222 |
| **Jane** | 0.111 | 0.310 | 0.000 |

Table 3 A centrality measures values for social network from Figure 1.

### 2.5.4   Degree Prestige

A degree prestige shows how popular is individual by counting how many direct connection is directed to this individual so degree prestige the same as indegree measure [49, 32].

### 2.5.5   Influence Domain

Influence domain for node *x* is number of nodes which can reach node *x* (there exists path to node *x*) [45].

### 2.5.6   Proximity Prestige

A proximity prestige is very similar to the closeness. It is the closeness multiply by the influence domain ($I_x$)

$$P_p(x) = \frac{I_x}{\sum_{\substack{y \neq x \\ y \in A}} c(x,y)} \qquad (7)$$

A proximity prestige is normalized using

$$P_p(x) = \frac{(I_x)^2}{(n-1) \cdot \sum_{\substack{y \neq x \\ y \in A}} c(x,y)} \qquad (8)$$

where *n* is the number of nodes in a network. [49, 32].

### 2.5.7 Rank Prestige

A rank prestige (also called a status prestige) of an actor A is a function of the prestige ranks others actors from a social network. If many individuals with a high rank value are in contact with one actor, then this actor has higher prestige than actors connected to individuals with lower rank value. "It's not what you know, but whom you know"[49].

## 2.6 Virtual Social Networks

A virtual social network is a type of a social network where the actors are connected, meet or cooperate through the Internet. Additionally, only a person can be an actor. Actors communicate and maintain their relationships using web services [50].

One of the first definition of virtual social network was proposed by Wasserman and Faust in [49] but more up-to-date definition can be found in [34]: *"A virtual social network VSN=(M,R) is the social network SN=(M,R) in which M is the finite set of non-anonymous internet user accounts – internet identities, called network members, that communicate with one another or participate in common activities provided by internet services. An asymmetric relationship $(m_i,m_j) \in R$, which links member $m_i \in M$ to member $m_j \in M$, exists if and only if there exists any communication from $m_i$ to $m_j$. The set of members M must not contain isolated members, i.e. $\forall m_i \in M \; \exists m_j \in M, i \neq j \; ((m_i,m_j) \in R \lor (m_i,m_j) \in R)$, card(M)>1."*

In spite of the fact that social networks on the Internet have already been investigated in many different contexts and many definitions were created, they are not consistent. Also, almost every researcher gives these networks differently name: supported social networks (CSSN) [51], web communities [17, 14], web-based social networks [18], virtual communities [1] or online social networks [16].

The term web communities was first used in 1998 [17] and 2000 [14] to describe the set of web pages which describes the same domain. According to Adamic and Adar every single web page must be linked with the physical individual to be treated as a node in the online social network. Therefore, they investigate the relation between users' homepages and based on this data create a virtual community. Furthermore, the similar social network can also be formed from an email communication system [1]. At the same time, a computer-supported social network described in [16, 51] appear when the computer network connects people or organizations. In the end, Golbeck claims that a web-based social network must fulfill the next criteria: users have to create their relationships with others, the system have to support connections and relationships creation, and this relationships must be visible and browsable [18]. Facebook, MySpace or Nasza-klasa are examples of dedicated social network systems which meet these conditions.

Because of the fact that the virtual social networks are subset of the social networks, all measures and methods used in social network analysis can be easy utilized in the virtual social networks.

Features which distinguish mark social network as virtual social networks are as follows [34]:

- Lack of physical contact – only by distance, even very long distances.
- Easy to break up, suspend contacts or relationships.
- The possibility of simultaneously communication with many members and the possibility of easy switches between different communication channels.
- Generally the lack of direct correlation between virtual member identity – internet identity and their identity in the real world. In VSN member can be different person than in real world.
- Quite easy to gather the data about communication or common activities and process this data.
- The lower reliability of the data about users and their activities available on the Internet. Users of Internet services relatively frequently provide fake personal data due to privacy concerns

Many different social networks can be extracted from services used by people. The most known social networks: set of people who are linked to one another by hyperlinks placed on their homepages [1], a customers who buy the same stuffs in the same e-commerce [42], people who date using an online dating system [7], a group of people who share information by utilizing shared bookmarking systems [40] such as del.icio.us., the company staff that communicates with one another via email [2, 47, 11, 55] More examples was enumerated below:

- social services – Facebook, Nasza-klasa.pl
- e-mail – Gmail, Yahoo!
- Instant messaging systems – MSN, ICQ, GG
- auction systems – eBay, Allegro
- e – commerce – Amazon, Merlin
- VoIP – Skype
- broadcasting systems – YouTube, Flickr
- telecommunication – Vodafone, Orange

Those services satisfy the humans basic needs of belonging to a social group. That is why they are so popular. They also provide simple ways of both expressing one's feelings or staying anonymous.

# 3 METHOD OF KEY USERS EXTRACTION

Social position is a social network analysis measure developed at Wrocław University of Technology. This measure can be used to calculate the importance of every single member of the network. Because of the fact that social position serves to estimate value of single user it was utilized in this master thesis to extract key users from the social network derived from the telecommunication data.

The importance of a user described by social position depends on the social positions of his/her close neighbourhood and the strength of relationship between user and his/her neighbour. More precisely user's social position is inherited from his neighbours which activity is directed to user and level of inheritance strictly depends on strength of this activity. The activity strength of one user absorbed by another is called commitment and almost always presented as weights of edges (Figure 3)[34].

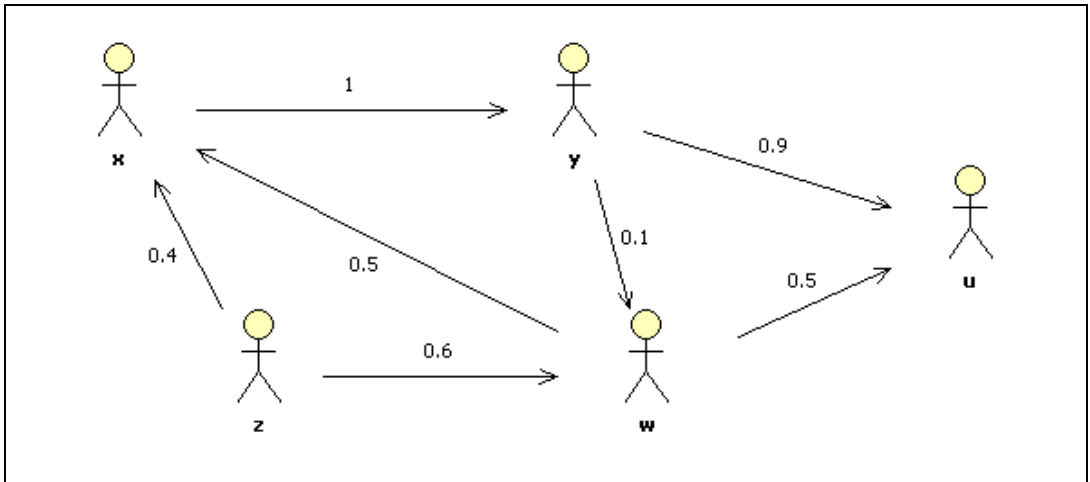

Figure 3 Social network with the assigned commitment values

## 3.1 Commitment Function Evaluation

To assess the strength of the relationship between two individuals $x$ and $y$ within the virtual social network the commitment function $C(y \to x)$ is used. It denotes the amount of the member $y$'s activity that person $y$ passes to member $x$ and is easily derived from relationship commitment function $C^{rel}(y \to x)$.

The commitment $C^{rel}(y \to x)$ of member $y$ within activity of their acquaintance $x$ is directly evaluated from source data as the normalized sum of all contacts, cooperation, and communications from $y$ to $x$ in relation to all activities of $y$:

$$C^{rel}(y \to x) = \begin{cases} \dfrac{A(y \to x)}{\sum_{x \in M} A(y \to x)}, & \text{when } \sum_{x \in M} A(y \to x) > 0 \\ 0, & \text{when } \sum_{x \in M} A(y \to x) = 0 \end{cases}, \qquad (9)$$

where:

$A(y \to x)$ – the function that denotes the activity of person $y$ directed to member $x$, e.g. number of emails sent by $y$ to $x$; $A(y \to x) \geq 0$;

$m$ – the number of people within the virtual social network.

Note that $A(y \to y)=0$, i.e. emails sent to themselves are excluded. Moreover, there may exist some inactive members $y$ in the network, for which $\sum_{x \in M} A(y \to x) = 0$ and in consequence $\sum_{x \in M} C^{rel}(y \to x) = 0$. Such inactive members $y$ are additionally proceeded at transformation from $C^{rel}(y \to x)$ to $C(y \to x)$ in the following way: if a member $y$ is not active to anybody, then some others members $x$ are active to $y$, since no isolated members are allowed in $VSN(M,R)$. In this case, the sum 1 is distributed equally among all $y$'s acquaintances – $x$ i.e. all values of $C(y \to x)$ (for more information see section 6.2 Social Position and Figure 4):

$$\forall (y \in M)$$
$$\sum_{z \in M} C^{rel}(y \to z) = 0 \Rightarrow \forall \left(x \in M : C^{rel}(x \to y) > 0\right) C(y \to x) = \frac{1}{card\left(\{x \in M : C^{rel}(x \to y) > 0\}\right)}. \quad (10)$$

The presence of the time is not considered in the formula (9). Similar approach is utilized by Valverde *et al.* where the strength of the relationships is established by the number of emails sent to a person in the group [48]. However the authors do not respect the general activity of the given individual. This general, local activity exists in the form of denominator in Eq. (9).

In another version of relationship commitment function $C^{rel}(y \to x)$ all member's activities are considered with respect to their time. The entire time from the first to the last activity of any member is divided into $k$ periods. For instance, a single period can be a month. Activities in each period are considered separately for each individual:

$$C^{rel}(y \to x) = \begin{cases} \dfrac{\sum_{i=0}^{k-1} (\lambda)^i \cdot A_i(y \to x)}{\sum_{x \in M} \sum_{i=0}^{k-1} (\lambda)^i \cdot A_i(y \to x)}, & \text{when } \sum_{x \in M} \sum_{i=0}^{k-1} (\lambda)^i \cdot A_i(y \to x) > 0 \\ 0, & \text{when } \sum_{x \in M} \sum_{i=0}^{k-1} (\lambda)^i \cdot A_i(y \to x) = 0 \end{cases}, \quad (11)$$

where:

$i$ – the index of the period: for the most recent period $i=0$, for the previous one: $i=1, \ldots$, for the most former $i=k-1$;

$A_i(y \to x)$ – the function that denotes the activity level of person $y$ directed to member $x$ in the $i$th time period, e.g. number of emails sent by $y$ to $x$ in the $i$th period;

$(\lambda)^i$ – the exponential function that denotes the weight of the $i$th time period, $\lambda \in (0;1]$;

$k$ – the number of time periods.

The activity of person $y$ is calculated in every time period and after that the appropriate weights are assigned to the particular time periods, using $(\lambda)^i$ factor. The most recent period $(\lambda)^i = \lambda^0 = 1$, for the previous one $(\lambda)^i = \lambda^1 = \lambda$ is not greater than 1, and for the most former period $(\lambda)^i = \lambda^{k-1}$ receives the smallest value. For example, if one year's data set is proceeded and a period is a month then $k=12$. For $\lambda=0.9$, the data from January is considered with the factor $0.9^{11}=0.31$, for February we have $0.9^{10}=0.35, \ldots$, for October $0.9^2=0.81$, for November – 0.9 and finally for December $0.9^0=1$. This in a sense is similar to an idea which was used in the personalized systems to weaken older activities of recent users [29].

One of the activity types is the communication via chat. In this case, $A_i(y \to x)$ is the number of chats that are common for *x* and *y* in the particular period *i*; and $\sum_{x \in M} A_i(y \to x)$ is the number of all chats in which *y* took part in the *i*th period. If person *y* had many common chats with *x* in comparison to the number of all *y*'s chats, then *x* has greater commitment within activities of *y*, i.e. $C^{rel}(y \to x)$ will have greater value and in consequence the social position of member *x* will grow.

Note that $C^{rel}(y \to x)$ will have value 1 when member *x* is the only interlocutor of person *y*.

However, not all of the elements can be calculated in such a simple way. Other activities are much more complex, e.g. comments on forums or blogs. Each forum consists of many threads where people can submit their comments. In this case, $A_i(y \to x)$ is the number of user *y*'s comments in the threads in which *x* has also commented, in period *i*, whereas sum $\sum_{x \in M} A_i(y \to x)$ is the total number of comments that have been made by all *x* who are *y*'s friends on these threads, at the same time.

---

**Commitment Evaluation Algorithm**

**Input:**
- *D* – data about communication, interaction or common activities between members *M* in the virtual social network *VSN*=(*M*,*R*).

**Output:**
- *C* – list that consists the commitment value for each ordered pair $(x_1, x_2) \in M$

1. begin
2. for (*each pair* $(x,y) \in M$) do
3.   *evaluate* $C^{rel}[x,y]$ *from D, e.g. using Eq. (9) or Eq. (10)*;
4. for (*each member* $x \in M$) do
5. begin
6.   commitment_of_x:=0;
7.   acquaintances_of_x:=0;
8.   for (*each member* $y \in M$) do
9.   begin
10.    commitment_of_x:=commitment_of_x+$C^{rel}[x,y]$;
11.    if ($C^{rel}[y,x]>0$) then
12.     acquaintances_of_x:=acquaintances_of_x+1;
13.   end;
14.   for (*each member* $y \in M$) do
15.    if ($C^{rel}[x,y]>0$) then
16.     $C[x,y]:=C^{rel}[x,y]$;
17.    else
18.     if (commitment_of_x=0 and $C^{rel}[y,x]>0$) then
19.      $C[x,y]:=1/$acquaintances_of_x;
20.     else
21.      $C[x,y]:=0$;
22. end;
23. end.

## 3.2 Social Position

Social position $SP(x)$ of member $x$ in social network $(A,C)$ is calculated by utilizing the values of social positions of all other network users and the level of their activities in relation to $x$. It is determined as follows:

$$SP(x) = (1 - \varepsilon) + \varepsilon \cdot \sum_{y \in M} SP(y) \cdot C(y \to x) \tag{12}$$

where:

$\varepsilon$ – the coefficient from the range (0;1).

$C(y \to x)$ – the commitment function which expresses the strength of the relation from $y$ to $x$.

The value of the constant $\varepsilon$ represents the openness of human social position on external influences, in other words high $\varepsilon$ means that the social position is highly influenced by others and low $\varepsilon$ means that the social position is more static and others influence is week.

Commitment function $C(y \to x)$ is a slightly modified version of relationship commitment $C^{rel}(y \to x)$. Function $C^{rel}(y \to x)$ describes the relationship data within the virtual social network $VSN(M,R)$.

Four important constraints regarding commitment function derived from the relationships $C^{rel}(y \to x)$ have to be fulfil [34]:

1. Relationship commitment function $C^{rel}(y \to x)$ is derived from the data describing relationships from $y$ to $x$ in $VSN(M,R)$, $x,y \in M$, $x \neq y$. If there exists the relationship $(y,x)$ $R$ the $C^{rel}(y \to x) > n$. If there is no relationship from $y$ to $x$, i.e. $(y,x) \notin R$ then $C^{rel}(y \to x) = 0$.

2. The value of relationship commitment is from the range [0;1]: $\forall (x, y \in M) C^{rel}(y \to x) \in [0;1]$

3. Relationship commitment function to itself equals 0: $\forall (y \in M) C^{rel}(y \to y) = 0$

4. If at least one relationship commitment from y is greater than 0, then the sum of all relationship commitments from y has to equal 1:

$$\forall (y \in M) \exists (x \in M) C^{rel}(y \to x) > 0 \Rightarrow \sum_{z \in M} C^{rel}(y \to z) = 1 \tag{13}$$

But condition 4 has to be satisfied by all network members $y$, not only those for whom $\exists (x \in M) C^{rel}(y \to x) > 0$, an additional condition has to be appended to the final commitment function $C(y \to x)$.

The new set of conditions for commitment function $C(y \to x)$ in $VSN(M,R)$ was presented below [34]:

1. Commitment function $C(y \to x)$ describes the strength of the relationship from $y$ to $x$ in $VSN(M,R)$, $x,y \in M$, $x \neq y$ and for that reason if $C^{rel}(y \to x) > 0$ then $C(y \to x) = C^{rel}(y \to x) = 0$, and $C^{rel}(y \to x)$ is the value of relationship commitment directly derived from the data about relationship (activities) from $y$ to $x$. If there is no relationship from $y$ to $x$ then $C^{rel}(y \to x) = C(y \to x) = 0$, except condition 5.

2. The value of commitment is from the range [0;1]: $\forall (x, y \in M) C(y \to x) \in [0;1]$.

3. Commitment function to itself equals 0: $\forall (y \in M) C(y \to y) = 0$.

4. The sum of all commitments has to equal 1, separately for each network member:

$$\forall (y \in M) \sum_{x \in M} C(y \to x) = 1 \quad (14)$$

5. If a member y is not active to anybody, then some others members x are active to y, since based on virtual social definition no isolated members are allowed in *VSN(M,R)* i.e. $\forall (y \in M) \exists (x \in M) C^{rel}(y \to x) = 0 \Rightarrow \sum_{z \in M} C^{rel}(y \to z) > 1$. In this case, to satisfy condition 4 (Eq. 11), the sum 1 is distributed equally among all y's acquaintances – x (Figure 4), i.e. all values of $C(y \to x)$ Eq. 10

The value of commitment function $C(y \to x)$ from *y* to *x* is usually derived from raw data about activity of member *y* directed to *x* or, in case of the total lack of *y*'s activity, as the equal potential contribution in activity.

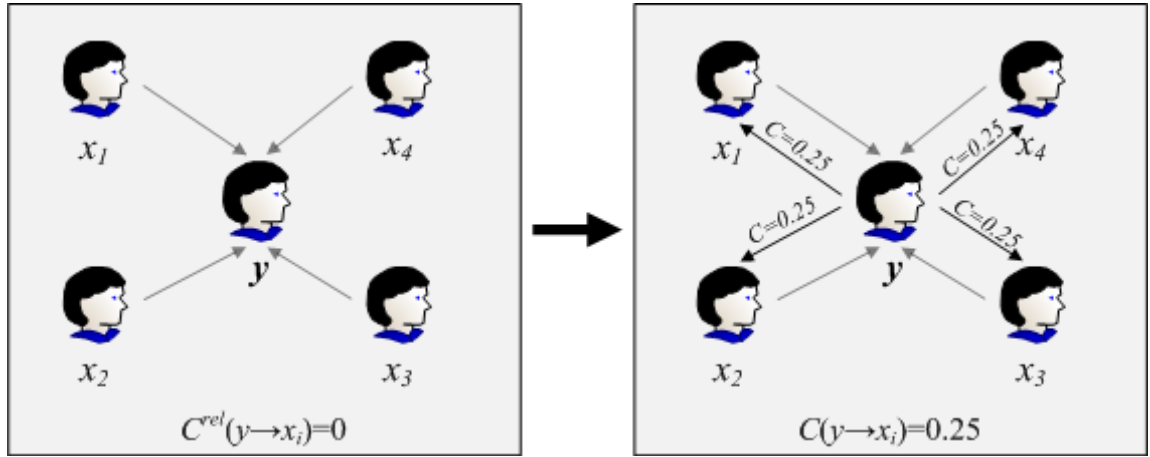

Figure 4 Distribution of the commitment for an inactive member *y* equally among all *y*'s acquaintances

Member *y* from Figure 4 has no connection to anybody within the network, but there are four other members (*x1, x2, x3, x4*) who are connected to user *y*. In such case the commitment function is equally distributed among all *y*'s contacts.

The virtual social network *VSN(M,R)* must not contain any isolated members (Definition of *VSN*). This restriction is derived from the lack of possibility to satisfy all enumerated above conditions for such members, especially condition 4 (Eq. 11) [34].

If member *y* is active to only one other member *x*, then $C(y \to x) = 1$ this is the consequence of the 4th constraint.

To satisfy the above requirements for the commitment function $C(y \to x)$ formula (12) can be expressed in a modified version. Social position function *SP(x)* of member *x* in *VSN=(M,R)* use only the values of social positions of direct member's *x* contacts as well as their activities in relation to *x* [34]:

$$SP(x) = (1 - \varepsilon) + \varepsilon \cdot \sum_{i=1}^{m_x} SP(y_i) \cdot C(y_i \to x) \quad (15)$$

where:

$y_i$ – *x*'s contacts, i.e. the members which relationship are directed to *x*: $C(y_i \to x) > 1$;

$m_x$ – the number of *x*'s contacts.

The reduction of element number in sum Eq. (15) compared to Eq. (12) can be important from the implementation point of view.

## 3.3 The SPIN Algorithm

The social position is calculated in the iterative way that means that the left side of Eq. (16) is the result of iteration while the right side is the input:

$$SP_{n+1}(x) = (1-\varepsilon) + \varepsilon \cdot \sum_{y \in M} SP_n(y) \cdot C(y \to x), \qquad (16)$$

where $SP_{n+1}(x)$ and $SP_n(x)$ is the social position of member $x$ after the $n+1$st and $n$th iteration, respectively.

In order to perform the first iteration, an initial value of social position $SP_0(x)$ for all $x \in M$ is needed:

$$SP_1(x) = (1-\varepsilon) + \varepsilon \cdot \sum_{y \in M} SP_0(y) \cdot C(y \to x). \qquad (17)$$

Since the algorithm is iterative, we also need to introduce a stop condition. For this purpose, a fixed precision coefficient $\tau$ is used. Thus, the calculation is stopped when the following criterion is met:

$$\forall (x \in M) \; |SP_n(x) - SP_{n-1}(x)| \leq \tau. \qquad (18)$$

Obviously, another version of the stop condition can be also applied, e.g.:

$$|SSP_n - SSP_{n-1}| \leq \tau,$$

where $SSP_n$ and $SSP_{n-1}$ is the sum of all social positions after the $n$th and $n$th iteration, respectively.

Based on Eq. (16), Eq. (17) and Eq. (18) we can develop the SPIN algorithm (*Social Position In the Network*). Three versions of this algorithm are proposed in this thesis, i.e. SPIN$^{node}$, SPIN$^{hybrid}$, and SPIN$^{edge}$. These algorithms differ in the implementation and in consequence their efficiency varies (see section 6.4 – Efficiency Tests).

All algorithms require the same set of input data and provide as the output the social position values for each network member and their ranking position regarding its social position as well as the number of iterations and time that was required to meet the stop condition that is one of the input parameters. Other input data that must be provided in order to evaluate the social position are: the list $C$ that consists the commitment value for each ordered pair $(x_1, x_2) \in M$, the initial social position for each member of the network, $\varepsilon$ coefficient from range [0,1].

### 3.3.1 SPIN$^{nodes}$

The first proposed algorithm SPIN$^{nodes}$ is the direct implementation of the social position concept. It is done without any optimization techniques. The name of the algorithm comes from the fact that all calculations are made form so called "node perspective", i.e. that the social position is calculated one by one for each network node – member.

First, two lists $SP_{prev}$ and $SP_{next}$ that contain the social position values are created. $SP_{prev}$ serves to store social positions from the previous iteration whereas in $SP_{next}$ the social positions calculated in the current iteration are stored. At the beginning, the initial social positions values $SP_0$ are assigned to the elements from $SP_{prev}$.

After that for each member $x$ from $M$ its $SP_{next}$ is set to 1-$\varepsilon$. Next, for each member $y$ from $M$ the value of commitment function $C(y \to x)$ is multiplied by $SP_{prev}[y]$ and by $\varepsilon$. The result of this operation is added to the current value of $x$'s social position that is stored in $SP_{next}[x]$. Finally, the values from $SP_{next}$ are assigned to $SP_{prev}$ and the iteration if finished. The next iteration is performed if the stop condition is not fulfilled. Otherwise the whole process is completed.

---

**The SPIN$^{nodes}$ algorithm**

**Input:**
- $C$ – list that consists the commitment value for each ordered pair $(x_1, x_2) \in M$
- $SP_0 = <SP_0(x_1), SP_0(x_2),…,SP_0(x_m)>$ – the vector of initial social positions, $m = \text{card}(M)$.
- $\varepsilon$ – coefficient from Eq. (9), $\varepsilon \in [0;1]$.
- $\tau$ – stop condition, i.e. the precision coefficient, e.g. $\tau = 0.00001$.

**Output:**
- $SP_{next} = <SP(x_1), SP(x_2),…,SP(x_m)>$ – the vector of final social positions.
- $R$ – the ranking of individuals from $M$.
- $n$ – the number of iterations.
- $t$ – processing time.

1. begin
2. $n := 0$;
3. $SP_{prev} := SP_0$;
4. repeat
5.   begin
6.   for (*each member x from M*) do
7.    begin
8.     $SP_{next}[x] := (1-\varepsilon)$;
9.     for (*each member y from M*) do
10.      $SP_{next}[x] := SP_{next}[x] + \varepsilon * SP_{prev}[y] * C[y,x]$;
11.    end;
12.   $SP_{prev} := SP_{next}$;
13.   $n := n+1$;
14.   end;
15. until *stop condition Eq. (18) is fulfilled for all members*;
16. *create ranking list R based on* $SP_{next}$;
17. end.

### 3.3.2 SPIN$^{edges}$

The second developed algorithm is called SPIN$^{edges}$ and its name comes from the fact that all calculations are made form so called "edge perspective", i.e. that the social position is calculated rather by taking into the consideration the edges and their weights (commitment functions assigned to the edges) then evaluating social position one by one for each network node – member.

First, the lists $SP$ that contains the initial social position values is created by assigning $SP_0$ to $SP$, i.e. that initially all $SP$s equal 0 for each network member.

After that for each edge $r(x,y)$ from the set $R$ of all edges in the network add to social position value of user $y$ ($SP(y)$) social position of $x$ ($SP(x)$) multiplied by the value of commitment function from user $x$ to $y$ ($C(x \rightarrow y)$). Next for each member of $M$ multiply the obtained social position of the given user by $\varepsilon$ and add the coefficient $1- \varepsilon$.

The next iteration is performed if the stop condition is not fulfilled. Otherwise the whole process is completed.

<div style="border:1px solid">

**The SPIN$^{edges}$ algorithm**

**Input:**
- $C$ – list that consists the commitment value for each ordered pair $(x_1, x_2) \in M$
- $SP_0 = <SP_0(x_1), SP_0(x_2), \ldots, SP_0(x_m)> = <0,0,\ldots,0>$ – the vector of initial social positions, $m = \text{card}(M)$.
- $\varepsilon$ – coefficient from Eq. (9), $\varepsilon \in [0;1]$.
- $\tau$ – stop condition, i.e. the precision coefficient, e.g. $\tau = 0.00001$.

**Output:**
- $SP = <SP(x_1), SP(x_2), \ldots, SP(x_m)>$ – the vector of final social positions.
- $R$ – the ranking of individuals from $M$.
- $n$ – the number of iterations.
- $t$ – processing time.

1. begin
2. $n := 0$;
3. $SP := SP_0$;
4. repeat
5.   begin
6.    for (*each edge r(x,y) from R*) do
7.     $SP[y] := SP[y] + SP[x]*C[x, y]$;
8.    for (*each member x from M*) do
9.     $SP[x] := (1 - \varepsilon) + \varepsilon * SP[x]$;
10.    $n := n+1$;
11.   end;
12. until *stop condition Eq. (18) is fulfilled for all members;*
13. *create ranking list R based on SP*;
14. end.

</div>

### 3.3.3 SPIN$^{hybrid}$

The third developed algorithm is called SPIN$^{hybrid}$ and it combined two previous approaches.

First, the lists *SP* that contains the initial social position values is created by assigning $SP_0$ to $SP$, i.e. that initially all *SP*s equal 0 for each network member.

After that all nodes of the network are divided into $m$ disjunctive subsets $\{s_1, s_2,\ldots, s_m\}$. Next steps are repeated until the stop condition is fulfilled. For each created subset $s_k$ the following action is performed: for each edge $r(x,y)$ in which $y$ belongs to subset $s_k$ add to social position value of $y$ ($SP(y)$) social position of $x$ ($SP(x)$) multiplied by the value of commitment function from user $x$ to $y$ ($C(x \rightarrow y)$). Next for each member of $M$ multiply the obtained social position of the given user by $\varepsilon$ and add the coefficient $1-\varepsilon$.

If the stop condition is fulfilled then the whole process is completed.

<div style="border:1px solid">

**The SPIN$^{hybrid}$ algorithm**

**Input:**
- $C$ – list that consists the commitment value for each ordered pair $(x_1, x_2) \in M$
- $SP_0 = <SP_0(x_1), SP_0(x_2), \ldots, SP_0(x_m)> = <0,0,\ldots,0>$ – the vector of initial social positions, $m = \text{card}(M)$.
- $m$ – the number of members processing in one step

</div>

- $\varepsilon$ – coefficient from Eq. (9), $\varepsilon \in [0;1]$.
- $\tau$ – stop condition, i.e. the precision coefficient, e.g. $\tau$:=0.00001.

**Output:**
- $SP=<SP(x_1),SP(x_2),…,SP(x_m)>$ – the vector of final social positions.
- $R$ – the ranking of individuals from $M$.
- $n$ – the number of iterations.
- $t$ – processing time.

```
1. begin
2.  n:=0;
3.  SP:=SP_0;
4.  divide the set M into m disjunctive subsets {s_1, s_2,…s_m}
5.  repeat
6.   begin
7.    for (each disjunctive subset s_k) do
8.     for (each edge r(x,y) where y is member of s_k) do
9.      SP[y]:=SP[y]+SP[x]*C[x,y];
10.   for (each member x from M) do
11.    SP[x]:=(1- ε) + ε *SP[x];
12.   n:=n+1;
13.  end;
14. until stop condition Eq. (18) is fulfilled for all members;
15. create ranking list R based on SP;
16. end.
```

# 4 SOCIAL NETWORK ANALYSIS PLATFORM

General concept of the Social Network Analysis Platform (SNAP) was created by the Social Network Group at Wrocław University of Technology. The aim was to join work of few students who were writing master thesis from the area of the computational social networks. Their shared some parts of their work. Thanks to that, work was easier and some SNAP parts were build and this parts can be used by others student in the future. The idea is to develop SNAP further by Social Network Group members and master thesis students.

## 4.1 Common Modules

Common Modules is a set of the SNAP's modules, which are developed by all students. Because of that this modules can be split among these students. Figure 5 shows all common modules and data flow between them. The grey modules were implemented by author of this master thesis so they was presented more precisely than others.

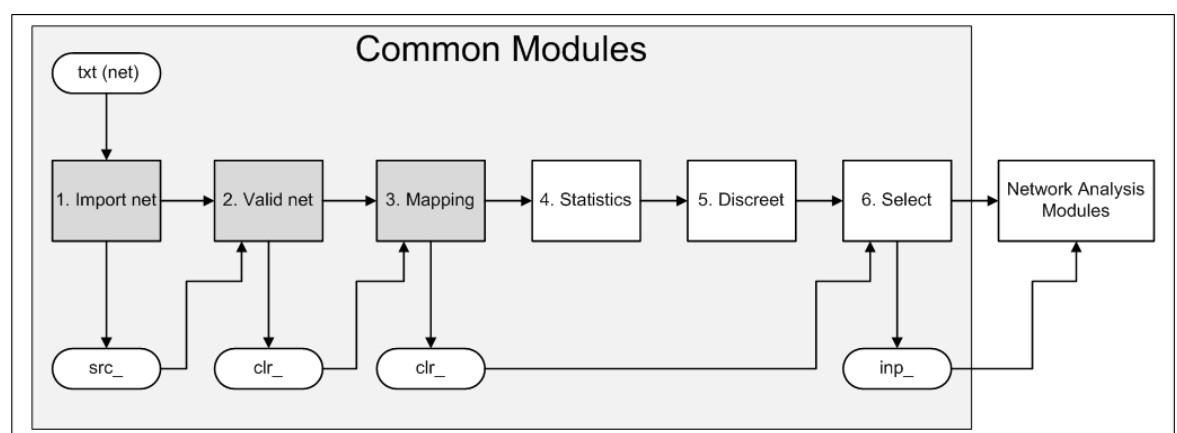

Figure 5. Common modules and a data flow between them. Modules marked on grey was made by this master thesis author

### 4.1.1    Import Net Module

This module is responsible for importing the data about the network from a text files to the SNAP's database. User provide 3 different files in special format. First file, Metadata, is mandatory and describes a network. File contains one row with 13 columns separated by semicolons:

1. snet_src_id – network id given by user

2. snet_name – network name given by user

3. snet_direct – flag which define if network is directed or not (1 – directed, 0 – not directed)

4. snet_weight_n – flag which define if nodes are weighted or not (1 – weighted, 0 – not weighted)

5. snet_weight_e – flag which define if edges are weighted or not (1 – weighted, 0 – not weighted)

6. snet_colour_n – flag which define if nodes are coloured or not (1 – coloured, 0 – not coloured)

7. snet_colour_e – flag which define if edges are coloured or not (1 – coloured, 0 – not coloured)

8. snet_layer_n – flag which define if nodes have layers or not (1 – have layers, 0 – don't have layers)
9. snet_layer_e – flag which define if edges have layers or not (1 – have layers, 0 – don't have layers)
10. snet_date – date of network creation or import given by user in format YYYY-MM-DD
11. snet_description – network description given by user
12. snet_gen_node – flag which define if nodes should be created (1 – should be, 0 – should not be). Flag only works when user doesn't provide Nodes file.
13. snet_seq_index – sequence index given by user

This row is split into columns and saved in src_net table (Figure 6) in plain text.

| src_net | src_node | src_node_err |
|---|---|---|
| snet_id | sn_id | err_id |
| snet_src_id | sn_net_id | sn_id |
| snet_name | sn_src_node | err_desc |
| snet_direct | sn_weight | |
| snet_weight_n | sn_colour | |
| snet_weight_e | sn_layer | |
| snet_colour_n | | |
| snet_colour_e | **src_edge** | **src_edge_err** |
| snet_layer_n | se_id | err_id |
| snet_layer_e | se_net_id | se_id |
| snet_date | se_src_from | err_desc |
| snet_description | se_src_to | |
| snet_gen_node | se_weight | |
| snet_seq_index | se_colour | |
| | se_layer | |

Figure 6 Tables used to store data by Import module

Second file, Edges, is also mandatory. File contains list of edges, one edge in one row. Each row has 5 columns separated by semicolons:

1. se_src_from – name of a start node
2. se_src_to – name of a end node
3. se_weight – edge's weight, if edges are not weighted field can be empty
4. se_colour – edge's colour, if edges are not coloured field can be empty
5. se_layer – edge's layer, if edges don't have layers field can be empty

Each row is split into columns and saved in src_edge table in plain text. All errors which occur during this process are stored in src_edge_err table (Figure 6).

The last file, Nodes, is not mandatory. It contains list of nodes, one node in each row. One row has 4 columns separated by semicolons:

1. sn_src_node – name of a node

2. sn_weight – node's weight, if nodes are not weighted field can be empty
3. sn_colour – node's colour, if nodes are not coloured field can be empty
4. sn_layer – node's layer, if nodes don't have layers field can be empty

Each row is split into columns and saved in src_node table in plain text. All errors which occur during this process are stored in src_node_err table (Figure 6). If user do not provide this file nodes can be automatically generate based on clr_edge (Figure 7) table during validation process.

### 4.1.2　　Valid Net Module

| clr_net | clr_node | clr_node_err |
|---|---|---|
| cnet_id | cn_id | err_id |
| cnet_src_id | cn_net_id | cn_id |
| cnet_name | cn_src_node | err_desc |
| cnet_direct | cn_weight | |
| cnet_weight_n | cn_colour | |
| cnet_weight_e | cn_layer | |
| cnet_colour_n | cn_weight_class | |
| cnet_colour_e | | |
| cnet_layer_n | | |
| cnet_layer_e | **clr_edge** | **clr_edge_err** |
| cnet_date | ce_id | err_id |
| cnet_description | ce_net_id | ce_id |
| cnet_gen_node | ce_src_from | err_desc |
| cnet_seq_index | ce_src_to | |
| | ce_from | |
| | ce_to | |
| | ce_weight | |
| | ce_weight_class | |
| | ce_colour | |
| | ce_layer | |

Figure 7 Tables used by Valid net module and Mapping Module

This module task is to map fields in the src_ tables to the clr_ tables and check the data against the graph theory. Whole process contains 3 steps:

1. Types mapping – in the first step every field form every row in the src_net, the src_edge and the src_node are mapped from a plain text to appropriate format i.e all weight values are mapped to numeric format. Mapped rows are saved in appropriate clr_ tables (Figure 7). All incorrect row are additionally saved in the clr_node_err or the clr_edge_err table and rapport to a user.

2. Checking against metadata – in this step all edges and nodes are checked against metadata i.e. if in metadata flag cnet_weight_e (flag has the same meaning as snet_weight_e) has value 1, then all the edges in clr_edge have to have weight values and if flag is 0 then edges cannot have weight values. All incorrect row are saved in the clr_node_err or the clr_edge_err table and rapport to a user.

3. Checking data against the graph theory – network has to fulfil three conditions to fulfil

graph theory constrains:

1) every triple (se_from, se_to, se_layer) in the clr_edge table has to be unique.

2) each ce_src_from and ce_src_to from the clr_edge table have to be in the clr_node table. One exception is when user did not provide Nodes file. If flag cnet_gen_node has value 1 then all missing nodes are automatically generated.

3) if edges have layers (flag cnet_layer_e set on 1) then each ce_src_from and ce_src_to from the clr_edge table have to be in the clr_node table and have to have appropriate layer in this table.

### 4.1.3  Mapping Module

Almost always name of the node is in plain text like an e-mail address or a user login. Processing such data is very difficult and inefficient. Mapping Module task is to map node's name from text to natural number. For each row in clr_edge table algorithm finds ce_src_from in clr_nodes table and saves its cn_id in ce_from column. The same procedure is executed for ce_src_to. After this process all clr_ tables are complete.

### 4.1.4  Others Modules from Common Modules

- Statistics Module – module which provides different information about network like number of edges, number of nodes, nodes degree, network degree and many others.

- Discreet Module – module which, when it is needed, converts continuous values to discreet values.

- Select Module – module responsible for choosing appropriate, defined by user, columns and rows and saving them in a inp_ tables. This tables is a start table for all Network Analysis Modules.

## 4.2 Network Analysis Modules

Network Analysis Modules are a set of a individual modules build by different authors depends of their needs. Each module contains an algorithm or set of the algorithms responsible for some part of social network analysis e.g. Social Position Module contains set of algorithms to count user social positions. This modules set will be strongly developed in future and allow to add many different modules to support social network analysis.

## 4.3 Social Position Module

Social Position module contains set of algorithms and functions to calculate users social positions and help to storage and maintain data which are being gathered during social positions calculation. The core of this module is implementation of three algorithms to calculate users social positions: $SPIN^{node}$, $SPIN^{hybrid}$, and $SPIN^{edges}$ (see Sec. 3.3 The SPIN Algorithm). All three algorithms, after each iteration, creates log file which contains all users, their social positions, duration of the whole iteration and durations of a particular iteration section. To calculate social position tables soc_pos_ are used (Figure 8 and Figure 9). Soc_pos_edge contains the edges and their weight, in this case weight is just a value of commitment function. Soc_pos_node contains results from every completed iteration so in this table, during iteration are data from previous iteration. To save partial results generated during iteration the soc_pos_node_tmp table is used. After iteration all records are just copied to the soc_pos_node.

![soc_pos tables](soc_pos_edge, soc_pos_node, soc_pos_node_tmp)

Figure 8 Tables used by Social Position Module to calculate users social positions and store results

Social Position Module also has additional function:

- to import data from the inp_ or the clr_ tables to the soc_pos_ tables
- to import social positions from log file to the soc_pos_node table in order to repeat some iteration.
- to setting new initial social position for every user
- to count iteration statistic – minimum, maximum and average social position, standard derivation, number of social position lower or equal than 1, greater than 1 and lower or equal 10, greater than 10 and lower or equal 100, greater than 100 and lower or equal 1000 and grater then 1000
- to compare two iteration using the Kendall's coefficient of concordance

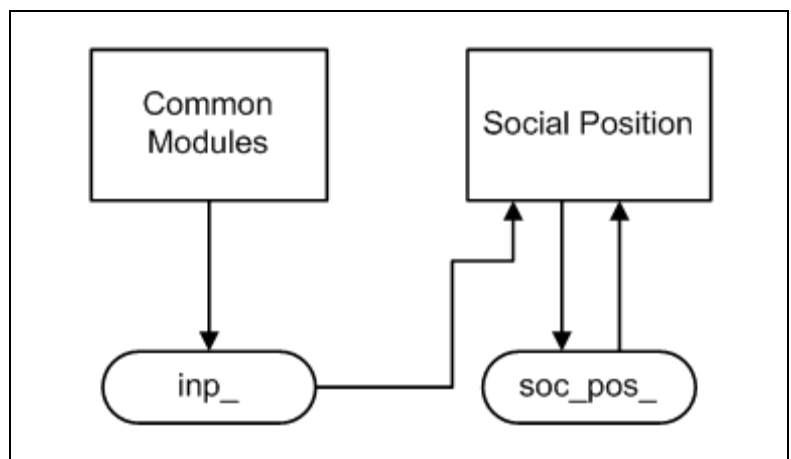

Figure 9 Social network module and it's data flow

# 5 TELECOMMUNICATION DATA

Telecommunication data has been delivered by Big European Telecomunication Provider[2] that is one of the largest United Kingdom telecommunication services provider.

## 5.1 Data Description

Data was delivered in btc.zip file and their size was 864 MB. After unpacking data was in plain text file (btc.txt) and their size was 4.83GB (4946MB). Text file contains 122.770.901 rows which describes phone calls in London during August 2005 (from 01.08.2008 to 31.08.2008). Each row describes one phone call and contains encrypted Caller telephone number, encrypted Receiver telephone number, date of the phone call in format YYMMDD, time of the phone call in format HHMMSS and duration of the phone call in seconds Figure 10.

---

1000008248,16505109637,050827,103157,33
1000121450,16505109637,050830,071805,33
1000807379,16505100096,050814,184053,1151

---

Figure 10 Example of telecommunication data. Caller telephone number (●), Receiver telephone number(●), date of the phone call(●), time of the phone call(●), duration of the phone call (●)

All callers are subscribers of Big European Telecomunication Provider but not all receivers are subscribers of Big European Telecomunication Provider.

Basic information about data:

- Provider – Big European Telecomunication Provider

- Size – 4,83GB (plain text)

- Time period from 2005-08-01 to 2005-08-31

- Number of phone calls – 122770901 (122.770.901)

- Number of phone calls which last zero seconds – 4509554 (4.509.554)

- Number of callers – 4455802 (4.455.802)

- Number of receivers – 8897388 (8.897.388)

- Number of different pair (caller-receiver) – 31159420 (31.159.420)

- Number of users which are both caller and receiver – 2442447 (2.442.447)

## 5.2 Data Pre-processing

At the beginning the data has to be imported to database. To import, storage and prepare data the tools provided by SAS Institute[3] were utilized.

---

[2] BT is one of the world's leading providers of communications solutions and services operating in 170 countries. Its principal activities include networked IT services, local, national and international telecommunications services, and higher-value broadband and internet products and services. More information on: http://www.bt.com/

[3] SAS is one of the largest software companies in the world. SAS tools to maintain and process huge amount of data. More information on: http://www.sas.com

1. Data was imported to table, AllData, which contains five columns Caller telephone number in String format, Receiver telephone number in String format, date of the phone call in date format DDMMYY, time of the phone call in time format HHMMSS and duration of the phone call in Numeric format. This table allowed to extract most of basic information which was listed above.
2. Next step was to extract from original data information about network user such as: user's telephone number, number of dialled calls, number of received call, duration of incoming calls and duration of outgoing calls, how many users called him and how many users he called. This data was inserted into table Users.
3. The third step was to extract from original data information about connection between pair (A,B) of users. Follow information was extracted A's telephone number, B's telephone number, number of calls from A to B, number of calls from B to A, duration of calls from A to B, duration of calls from B to A. This data was inserted into table Connections. Number of connection was 31159420.

After investigating tables Users and Connections it appears that some users have calls which duration time is 0,1 or 2 seconds. It was assumed that most of this users are automats which makes calls and according to the social network definition an automat cannot be a member of a social network so calls which last less than 3 seconds have to be removed. Of course some automats calls has left and maybe a few calls made by human was also removed, but for sure most of automats calls was deleted .

9.829.459 phone calls was removed from AllData and new table was saved as AllData_2. After that tables Users and Connections was calculated once again (For new table size see Table 4).

| Table | Description | Size [GB] | No. of rows |
|---|---|---|---|
| **AllData** | Complete data after import | 4.6 | 122.770.901 |
| **Users** | List of users extracted from AllData | 0,926 | 10.910.743 |
| **Connections** | List of connections extracted from AllData | 2.6 | 31.159.420 |
| **AllData_2** | All calls which duration was longer than 2 seconds | 4.3 | 112.941.442 |
| **Users_2** | List of users extracted from AllData_2 | 0.897 | 10.580.206 |
| **Connections_2** | List of connections extracted from AllData_2 | 2.5 | 30.231.102 |
| **AllData_2_net** | All calls which duration was longer than 2 seconds and both caller and receiver belongs to BT | 2.2 | 58.979.091 |
| **Users_2_net** | List of users extracted from AllData_2_net | 0.378 | 4.455.802 |
| **Connections_2_net** | List of connections extracted from AllData_2_net | 1.2 | 17.063.810 |

Table 4 Basic information about tables created during social networks creation.

Despite having smaller tables witch lesser amount of data the density of network which can be created from this data is to small (about $5,4 \cdot 10^{-7}$). Moreover only information about BT subscribers can be obtained. Because of this two fact it was decided to cut AllData_2 down and remove all phone calls directed to users which were not BT subscribers. Results was saved in AllData_2_net and once again table Users and Connections was calculated (Table 4). Thanks to this data cut, the density, of network which can be created from this data, increased (about 17,2•10$^{-7}$) and data about all network users can be delivered by BT.

Based on tables Users_2_net and Connections_2_net social network was built with two different commitment function.

First commitment function base on number of phone calls and can be expressed as:

$$C_N(A \rightarrow B) = \frac{N(A \rightarrow B)}{N(A)} \quad (19)$$

where :

N(A→B) – number of calls which user A made to user B

N(A) – number of A's calls.

Second commitment function base on duration of phone calls and can be expressed as:

$$C_D(A \rightarrow B) = \frac{T(A \rightarrow B)}{T(A)} \quad (20)$$

where :

T(A→B) – sum of durations of calls made by user A to user B

T(A) – sum of duration of all A's calls.

Using this two commitment functions and tables Users_2_net and Connections_2_net 3 files was prepared:

1. List of 4.455.802 users one in each row. Size 84.6MB
2. List of 17.063.810 connections witch first commitment function. One connection in each row. Each row contain UserA;UserB;CN(A→B). Size 621.2MB
3. List of 17.063.810 connections witch second commitment function. One connection in each row. Each row contain UserA;UserB;CD(A→B). Size 628.6MB

Data in file was ready to use by SNAP and perform experiments.

# 6 EXPERIMENTS

In Section 3 the mathematical basis and justification of commitment function and social position method were presented. The answers for research question 3-6 by were provided by investigating the features of social position measure, such as average, minimum and maximum value of SP, distribution of its values, etc., and by analysing the influence of ε coefficient on the social position value and its characteristics i.e. the mean value of SP, the distribution of SP, minimum and maximum values of SP for each ε and the ranking itself by designing and executing an experiment. Experiment is understood here (compared to frequently used framework [52]) as a systematic investigation of a certain phenomenon using a specific method in a specific context, where the phenomenon is the social position measure, method is a set of tests and the context is Big European Telecomunication Provider telecommunication data.

In order to answer research question no. 7, the social position measure was compared with indegree (IC) and outdegree (OC) centrality. Also to complete the answer for research question 6 and 7 the additional efficiency test were performed.

The experiment was chosen as the most suitable research method for purposes stated above because such method is often used by other researchers from social network field [1, 30, 31, 32, 34, 36, 42, 47]. Additionally due to security and users protection issues, the telecommunication data presented in this thesis was never analysed from social position point of view so the experiment was necessary to better understood the data and check if the proposed method is suitable for such data.. Computational environment was: the PC class computer with Windows XP (SP3), 2GB RAM memory, 7200 rpm and 400GB hard drive, and two core processor Intel Pentium D 3GHz.

## 6.1 The Calculation of Social Position

In order to calculate social position of users two variants of commitment function were utilized. First one takes into consideration number of phone calls – $C_N$ whereas the second one the duration of the phone calls – $C_d$. The former one was calculated in the following way:

$$C_N(y \to x) = \begin{cases} \dfrac{N(y \to x)}{\sum_{x \in M} N(y \to x)}, & \text{when } \sum_{x \in M} N(y \to x) > 0 \\ 0, & \text{when } \sum_{x \in M} N(y \to x) = 0 \end{cases}, \quad (21)$$

where:
$n(y \to x)$ – the number of phone calls that user $y$ made to user $x$;

$\sum_{x \in M} n(y \to x)$ – the number of all phone calls made by $y$ to all members of the social network $M$.

The value of latter commitment function, i.e. $C_T$ was evaluated as follows:

$$C_T(y \to x) = \begin{cases} \dfrac{T(y \to x)}{\sum_{x \in M} T(y \to x)}, & \text{when } \sum_{x \in M} T(y \to x) > 0 \\ 0, & \text{when } \sum_{x \in M} T(y \to x) = 0 \end{cases}, \qquad (22)$$

where:

$T(y \to x)$ – the duration of all phone calls that user y made to user x;

$\sum_{x \in M} T(y \to x)$ – the duration of all phone calls made by y to all members of the social network M.

The calculations were performed for all network members and in these calculations both described above commitment function were utilized. The value of $\varepsilon$ coefficient was established as from the set {0.1, 0.2, 0.3, 0.4, 0.5, 0.6, 0.7, 0.8, 0.9}. The SPIN$^{edge}$ algorithm was used in this part of experimental work.

| User ID | $\varepsilon=0.1$ | $\varepsilon=0.2$ | $\varepsilon=0.3$ | $\varepsilon=0.4$ | $\varepsilon=0.5$ | $\varepsilon=0.6$ | $\varepsilon=0.7$ | $\varepsilon=0.8$ | $\varepsilon=0.9$ |
|---|---|---|---|---|---|---|---|---|---|
| 16501077425 | 1 | 1 | 1 | 1 | 1 | 1 | 1 | 1 | 1 |
| 16501976647 | 2 | 2 | 2 | 2 | 2 | 2 | 2 | 2 | 3 |
| 16502258847 | 3 | 3 | 3 | 3 | 3 | 3 | 3 | 4 | 7 |
| 16501078947 | 4 | 4 | 4 | 4 | 4 | 4 | 4 | 3 | 5 |
| 16500608536 | 5 | 5 | 5 | 5 | 5 | 6 | 6 | 7 | 8 |
| 16501993971 | 6 | 6 | 6 | 7 | 7 | 7 | 8 | 9 | 10 |
| 16500981881 | 7 | 7 | 8 | 8 | 8 | 8 | 10 | 12 | 13 |
| 16500732058 | 8 | 8 | 7 | 6 | 6 | 5 | 5 | 6 | 6 |
| 16506564091 | 9 | 9 | 9 | 9 | 9 | 9 | 9 | 11 | 12 |
| 16501084780 | 10 | 10 | 10 | 11 | 13 | 14 | 15 | 15 | 16 |
| 16504555236 | 11 | 11 | 12 | 13 | 14 | 15 | 16 | 16 | 17 |
| 16506579643 | 12 | 12 | 13 | 12 | 12 | 13 | 14 | 14 | 14 |
| 16500983081 | 13 | 13 | 11 | 10 | 10 | 11 | 11 | 10 | 9 |
| 16504008948 | 14 | 14 | 15 | 15 | 16 | 16 | 17 | 19 | 20 |
| 16506320736 | 15 | 15 | 14 | 14 | 11 | 12 | 12 | 13 | 11 |

Table 5 The ranking of *SP* calculated based on the number of made phone calls for the first 15 users depending on $\varepsilon$ value

The analysis of the outcomes of the calculations revealed that the rankings of social position values calculated based on both number of made phone calls and overall duration of these phone calls are almost the same for each value of *ε* (Table 5, Table 6). The rankings for the first fifteen users are presented in Table 5 and Table 6. It shows that identified key users within the given telecommunication network are the same regardless the analyzed features of communication, i.e. number of phone calls as well as duration of them.

| User ID | $\varepsilon=0.1$ | $\varepsilon=0.2$ | $\varepsilon=0.3$ | $\varepsilon=0.4$ | $\varepsilon=0.5$ | $\varepsilon=0.6$ | $\varepsilon=0.7$ | $\varepsilon=0.8$ | $\varepsilon=0.9$ |
|---|---|---|---|---|---|---|---|---|---|
| 16501077425 | 1 | 1 | 1 | 1 | 1 | 1 | 1 | 1 | 1 |
| 16501976647 | 2 | 2 | 2 | 2 | 2 | 2 | 2 | 2 | 3 |
| 16502258847 | 3 | 3 | 3 | 3 | 3 | 3 | 3 | 4 | 6 |
| 16501078947 | 4 | 4 | 4 | 4 | 4 | 4 | 4 | 3 | 5 |
| 16500608536 | 5 | 5 | 5 | 5 | 5 | 6 | 6 | 7 | 8 |

| User ID | ε=0.1 | ε=0.2 | ε=0.3 | ε=0.4 | ε=0.5 | ε=0.6 | ε=0.7 | ε=0.8 | ε=0.9 |
|---|---|---|---|---|---|---|---|---|---|
| 16501993971 | 6 | 6 | 7 | 7 | 7 | 7 | 9 | 10 | 11 |
| 16500981881 | 7 | 8 | 8 | 8 | 8 | 9 | 10 | 12 | 14 |
| 16500732058 | 8 | 7 | 6 | 6 | 6 | 5 | 5 | 6 | 7 |
| 16506564091 | 9 | 9 | 9 | 9 | 9 | 10 | 11 | 11 | 13 |
| 16500983081 | 10 | 10 | 10 | 10 | 10 | 8 | 8 | 9 | 9 |
| 16501084780 | 11 | 11 | 11 | 11 | 11 | 14 | 14 | 15 | 17 |
| 16504555236 | 12 | 12 | 12 | 13 | 14 | 15 | 15 | 16 | 19 |
| 16506579643 | 13 | 13 | 13 | 12 | 13 | 13 | 13 | 14 | 16 |
| 16504008948 | 14 | 14 | 14 | 15 | 16 | 16 | 16 | 18 | 23 |
| 16506320736 | 15 | 15 | 15 | 14 | 12 | 12 | 12 | 13 | 12 |

Table 6 The ranking of *SP* calculated based on the duration of made phone calls for the first 15 users depending on $\varepsilon$ value

## 6.2 Distribution of Social Position

The important information about the values of social position provide the mean value of social position as well as the standard deviation of *SP* value. As in the previous section, all calculations were performed for all network members and also in this case two types of commitment function were utilized. One of them was evaluated based on the duration of the phone calls and the other one based on the number of phone calls.

The average values of social position and standard deviation of social position for different $\varepsilon$ values for BT phone calls where the commitment function was assessed based on the duration of phone calls are presented in Figure 11 and Table 7 whereas the same parameters for BT phone calls where the commitment function was assessed based on the number of phone calls are presented in Figure 12 and Table 8.

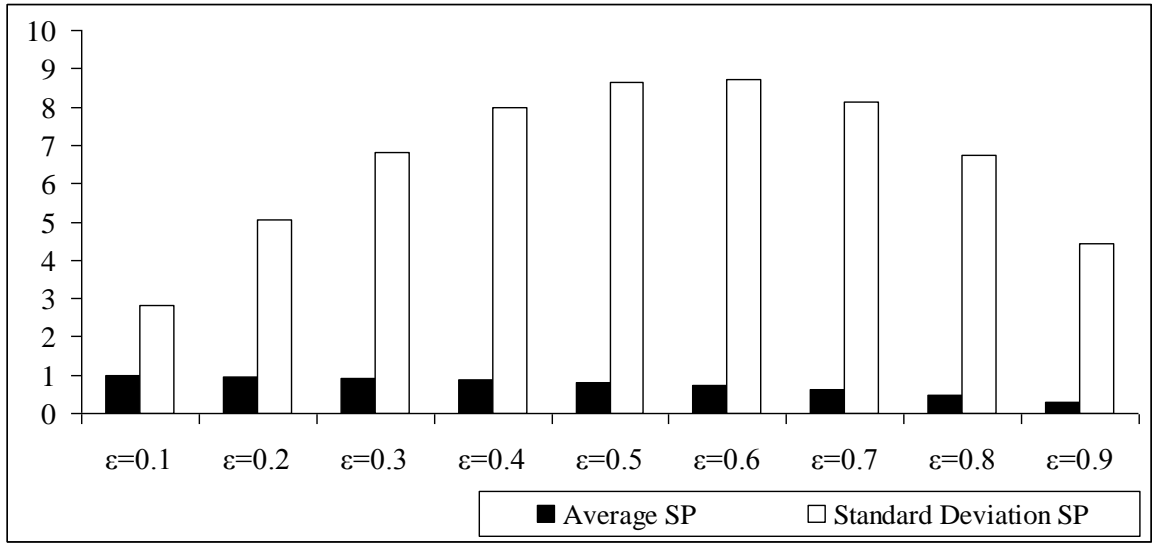

Figure 11 Average *SP*, its standard deviation in the BT dataset in relation to $\varepsilon$ where the commitment function was assessed based on the duration of the phone calls

The average social position does not depend on the value of $\varepsilon$ and it is convergent to 1. The values of average *SP* in Table 7 and Table 8 are the values after six iterations.

Although in Table 7 and Table 8 the average values of SP do not equal 1 it is convergent to 1 and can be formally proved however it is out of scope of this research. Note that the average social position value is convergent to 1 faster for smaller $\varepsilon$ values.

|  | $\varepsilon$=0.1 | $\varepsilon$=0.2 | $\varepsilon$=0.3 | $\varepsilon$=0.4 | $\varepsilon$=0.5 | $\varepsilon$=0.6 | $\varepsilon$=0.7 | $\varepsilon$=0.8 | $\varepsilon$=0.9 |
|---|---|---|---|---|---|---|---|---|---|
| Average *SP* | 0.98 | 0.96 | 0.92 | 0.87 | 0.80 | 0.71 | 0.61 | 0.47 | 0.31 |
| Std. Dev. *SP* | 2.82 | 5.07 | 6.80 | 8.00 | 8.65 | 8.72 | 8.12 | 6.73 | 4.44 |
| Min *SP* value | 0.90 | 0.80 | 0.70 | 0.60 | 0.50 | 0.40 | 0.30 | 0.20 | 0.10 |
| Max *SP* value | 4518.2 | 8148.7 | 10942.0 | 12907.7 | 14011.2 | 14164.7 | 13212.1 | 10910.8 | 6907.1 |

Table 7 Average *SP*, its standard deviation, minimum and maximum value of *SP* in the BT dataset in relation to $\varepsilon$ where the commitment function was assessed based on the duration of the phone calls

On the other hand, the standard deviation substantially differs depending on the value of coefficient $\varepsilon$. Overall, the greater $\varepsilon$, the bigger standard deviation. It shows that for greater $\varepsilon$ the value of the distance between the members' node positions increases. Moreover, if $\varepsilon$ is greater, the distance between the minimum and maximum social position within community increases and in consequence the standard deviation increases.

However, for both types of commitment functions the standard deviation of *SP* value decreases for *ε* greater than 0.7. In this situation it is caused by the fact that due to efficiency limits only six iterations were performed. The research performed on different datasets revealed that the standard deviation increases for greater *ε* [31]. This fact can be observed when the mean value of *SP* equals 1. So in case of BT dataset more iterations is required and it will be the goal of future research.

The minimum value of social position always equals (1- *ε*) for user who communicates with other people but with whom nobody communicates.

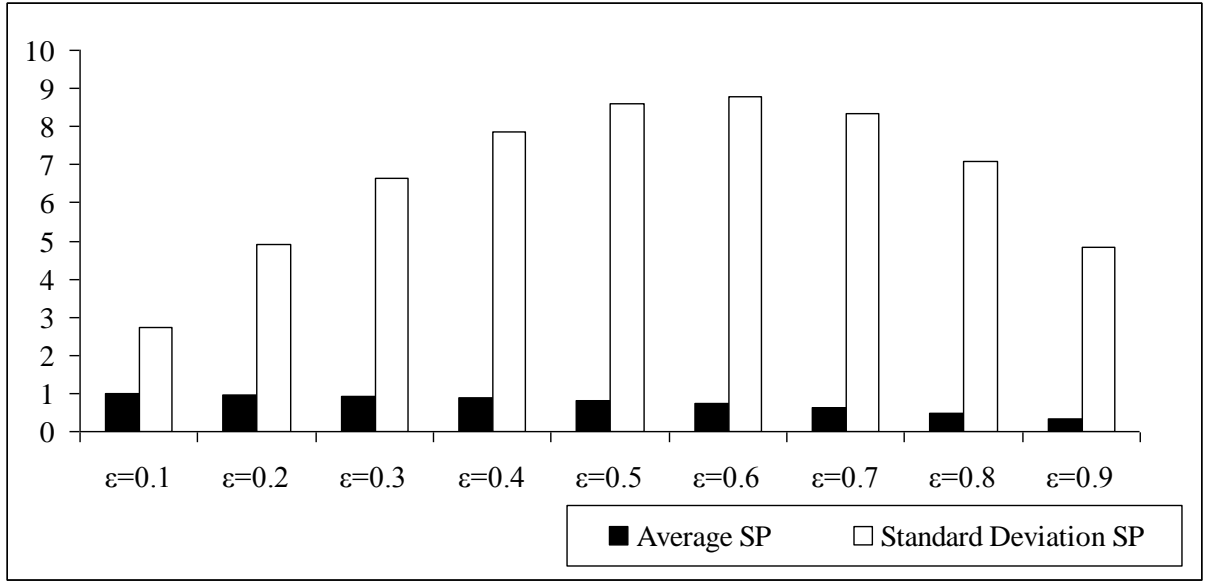

Figure 12 Average *SP*, its standard deviation in the BT dataset in relation to *ε* where the commitment function was assessed based on the number of the phone calls

|  | ε=0.1 | ε=0.2 | ε=0.3 | ε=0.4 | ε=0.5 | ε=0.6 | ε=0.7 | ε=0.8 | ε=0.9 |
|---|---|---|---|---|---|---|---|---|---|
| Average *SP* | 0.98 | 0.96 | 0.92 | 0.87 | 0.81 | 0.73 | 0.62 | 0.49 | 0.33 |
| Std. Dev. *SP* | 2.72 | 4.93 | 6.64 | 7.87 | 8.60 | 8.79 | 8.33 | 7.07 | 4.83 |
| Min *SP* value | 0.90 | 0.80 | 0.70 | 0.60 | 0.50 | 0.40 | 0.30 | 0.20 | 0.10 |
| Max *SP* value | 4331.7 | 7853.0 | 10634.3 | 12696.0 | 14001.6 | 14442.5 | 13813.6 | 11781.1 | 7841.4 |

Table 8 Average *SP*, its standard deviation, minimum and maximum value of *SP* in the BT dataset in relation to *ε* where the commitment function was assessed based on the number of the phone calls

Another feature of the social position measure is the distribution of its values (Table 9, Table 10, Table 11, Table 12). First the social position values were divided into five classes. The ranges of the classes are as follows: *SP*≤1, 1<*SP*<10, 10≤*SP*<100, 100≤*SP*<1000, and *SP*≥1000.

|  | ε=0.1 | ε=0.2 | ε=0.3 | ε=0.4 | ε=0.5 | ε=0.6 | ε=0.7 | ε=0.8 | ε=0.9 |
|---|---|---|---|---|---|---|---|---|---|
| SP<=1 | 4227986 | 4208573 | 4191595 | 4179301 | 4173238 | 4177383 | 4188601 | 4211780 | 4259114 |
| 1<SP<10 | 222593 | 237769 | 251638 | 261481 | 265961 | 261446 | 251217 | 230607 | 187665 |
| 10<=SP<100 | 5071 | 9044 | 11935 | 14182 | 15661 | 16015 | 15121 | 12779 | 8714 |
| 100<=SP<1000 | 149 | 408 | 621 | 822 | 923 | 939 | 846 | 624 | 302 |
| SP>=1000 | 3 | 8 | 13 | 16 | 19 | 19 | 17 | 12 | 7 |

Table 9 The number of users with SP belongs to the predefined classes within the BT dataset in relation to ε where the commitment function was assessed based on the duration of the phone calls

|  | ε=0.1 | ε=0.2 | ε=0.3 | ε=0.4 | ε=0.5 | ε=0.6 | ε=0.7 | ε=0.8 | ε=0.9 |
|---|---|---|---|---|---|---|---|---|---|
| SP<=1 | 94.887 | 94.452 | 94.070 | 93.795 | 93.659 | 93.752 | 94.003 | 94.523 | 95.586 |
| 1<SP<10 | 4.996 | 5.336 | 5.647 | 5.868 | 5.969 | 5.868 | 5.638 | 5.175 | 4.212 |
| 10<=SP<100 | 0.114 | 0.203 | 0.268 | 0.318 | 0.351 | 0.359 | 0.339 | 0.287 | 0.196 |
| 100<=SP<1000 | 0.003 | 0.009 | 0.014 | 0.018 | 0.021 | 0.021 | 0.019 | 0.014 | 0.007 |
| SP>=1000 | 0.0001 | 0.0002 | 0.0003 | 0.0004 | 0.0004 | 0.0004 | 0.0004 | 0.0003 | 0.0002 |

Table 10 The percentage of users with SP belongs to the predefined classes within the BT dataset in relation to ε where the commitment function was assessed based on the duration of the phone calls

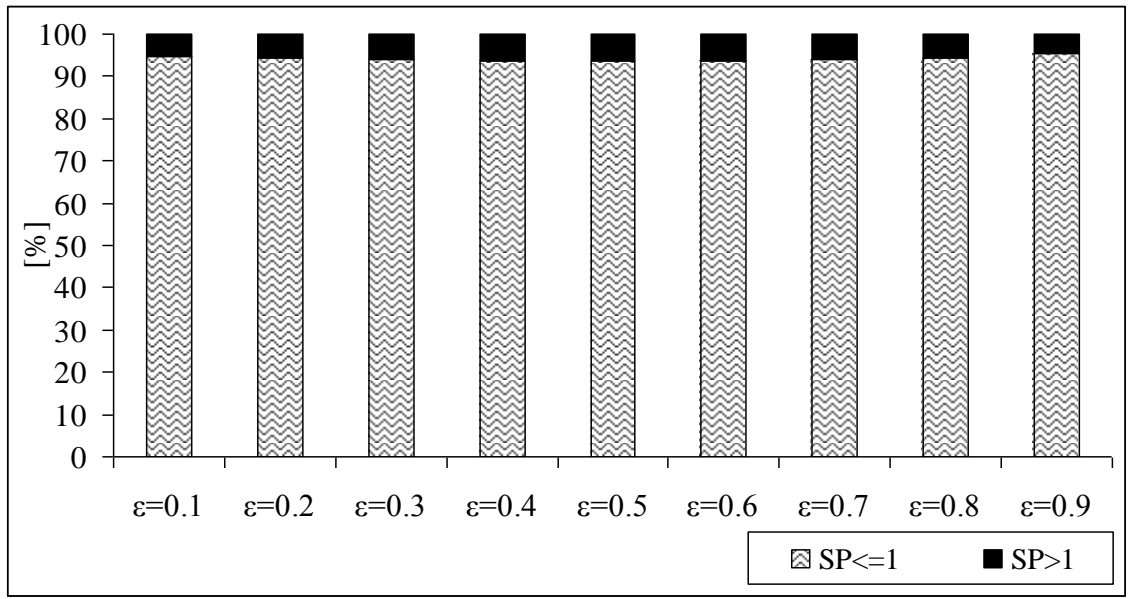

Figure 13 The percentage of users with *SP* belongs to the predefined classes within the BT dataset in relation to *ε* where the commitment function was assessed based on the duration of the phone calls

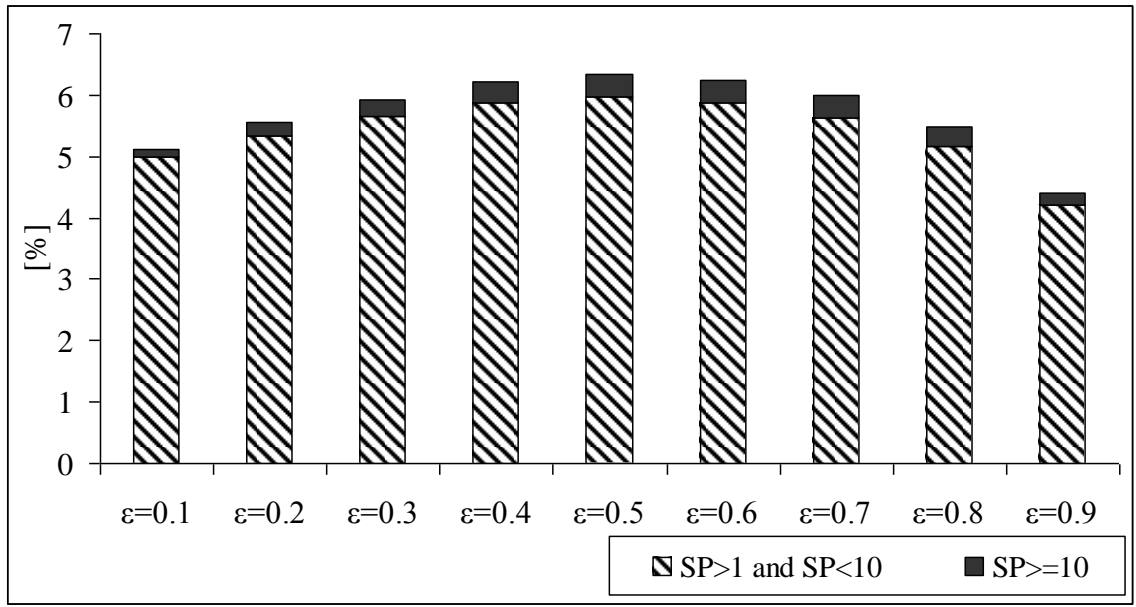

Figure 14 The percentage of users with *SP* belongs to the predefined classes within the BT dataset in relation to $\varepsilon$ where the commitment function was assessed based on the duration of the phone calls

It can be easily noticed that the vast number of users have obtained the social position measure lower than 1. Moreover, it does not depend on the way of commitment function evaluation (see Figure 13 and Figure 15). For both commitment functions over 93% of users have the social position lower than 1 regardless $\varepsilon$ value. Moreover, only around 5% of users obtained the *SP* value between 1 and 10 (Figure 14 and Figure 16). Finally, only around 1% of users have the social position greater than 10.

More detail analysis revealed that although the maximum value of social position is greater than 4,000 for each $\varepsilon$ value, there are only few users who obtained social position value greater than 1,000. For example for $\varepsilon=0.5$, where the commitment function was assessed based on the number of the phone calls, the maximum value of *SP* equals 14,001.66 while there are only 19 users who have the *SP* greater than 1,000 (Table 11), and it is only 0.0004% of the whole population. The analogous situations appear for other $\varepsilon$ values and when commitment function calculated based on the duration of phone calls.

This analysis confirms that node position can be the good measure to extract key users in the social network [30].

|  | $\varepsilon=0.1$ | $\varepsilon=0.2$ | $\varepsilon=0.3$ | $\varepsilon=0.4$ | $\varepsilon=0.5$ | $\varepsilon=0.6$ | $\varepsilon=0.7$ | $\varepsilon=0.8$ | $\varepsilon=0.9$ |
|---|---|---|---|---|---|---|---|---|---|
| SP<=1 | 4229033 | 4212494 | 4196784 | 4184520 | 4178316 | 4180957 | 4190979 | 4213292 | 4259069 |
| 1<SP<10 | 221442 | 233585 | 246010 | 255746 | 260331 | 257154 | 247970 | 228020 | 186467 |
| 10<=SP<100 | 5183 | 9304 | 12363 | 14675 | 16178 | 16683 | 15943 | 13797 | 9910 |
| 100<=SP<1000 | 142 | 412 | 632 | 846 | 958 | 989 | 894 | 680 | 349 |
| SP>=1000 | 2 | 7 | 13 | 15 | 19 | 19 | 16 | 13 | 7 |

Table 11 The number of users with SP belongs to the predefined classes within the BT dataset in relation to ε where the commitment function was assessed based on the number of the phone calls

|  | ε=0.1 | ε=0.2 | ε=0.3 | ε=0.4 | ε=0.5 | ε=0.6 | ε=0.7 | ε=0.8 | ε=0.9 |
|---|---|---|---|---|---|---|---|---|---|
| SP<=1 | 94.911 | 94.540 | 94.187 | 93.912 | 93.772 | 93.832 | 94.057 | 94.557 | 95.585 |
| 1<SP<10 | 4.970 | 5.242 | 5.521 | 5.740 | 5.843 | 5.771 | 5.565 | 5.117 | 4.185 |
| 10<=SP<100 | 0.116 | 0.209 | 0.277 | 0.329 | 0.363 | 0.374 | 0.358 | 0.310 | 0.222 |
| 100<=SP<1000 | 0.003 | 0.009 | 0.014 | 0.019 | 0.022 | 0.022 | 0.020 | 0.015 | 0.008 |
| SP>=1000 | 0.0000 | 0.0002 | 0.0003 | 0.0003 | 0.0004 | 0.0004 | 0.0004 | 0.0003 | 0.0002 |

Table 12 The number of users with SP belongs to the predefined classes within the BT dataset in relation to ε where the commitment function was assessed based on the number of the phone calls

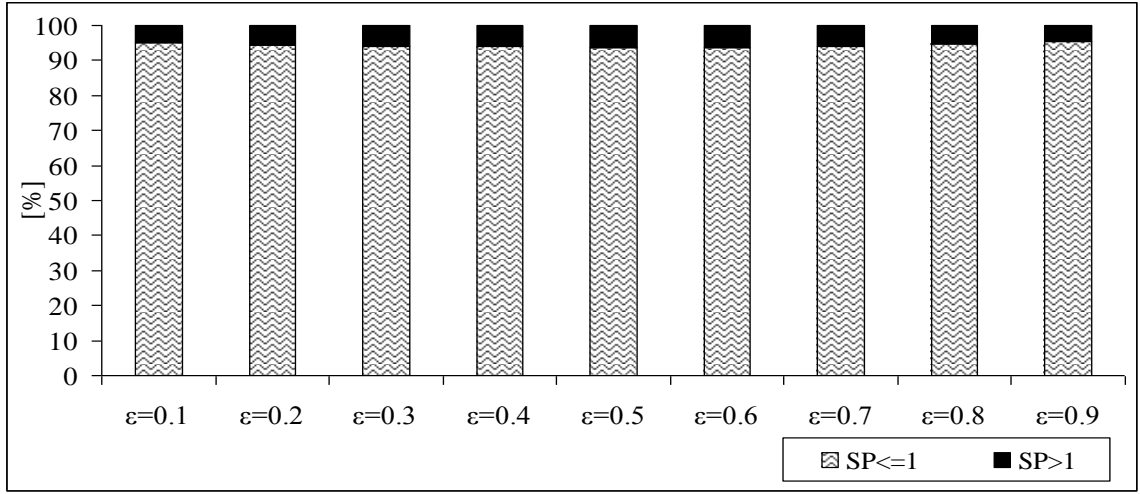

Figure 15 The percentage of users with *SP* belongs to the predefined classes within the BT dataset in relation to *ε* where the commitment function was assessed based on the number of the phone calls

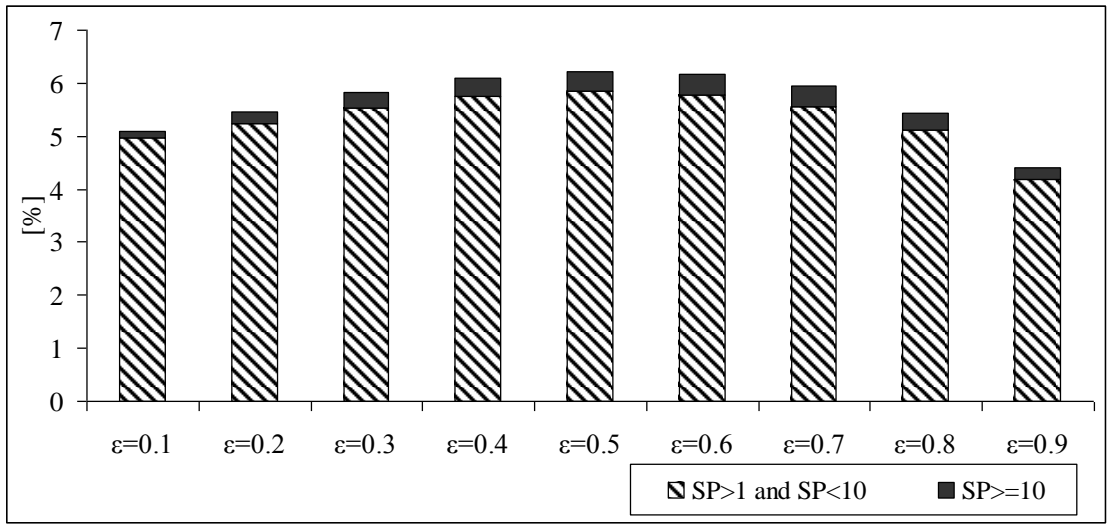

Figure 16 The percentage of users with *SP* belongs to the predefined classes within the BT dataset in relation to *ε* where the commitment function was assessed based on the number of the phone calls

In order to show the distribution of the social position values the charts that show the social position values for different ε values for the first 30,000 of users is presented in Figure 17 (where the commitment function was assessed based on the number of the phone calls) and Figure 19 (where the commitment function was assessed based on the duration of the phone calls).

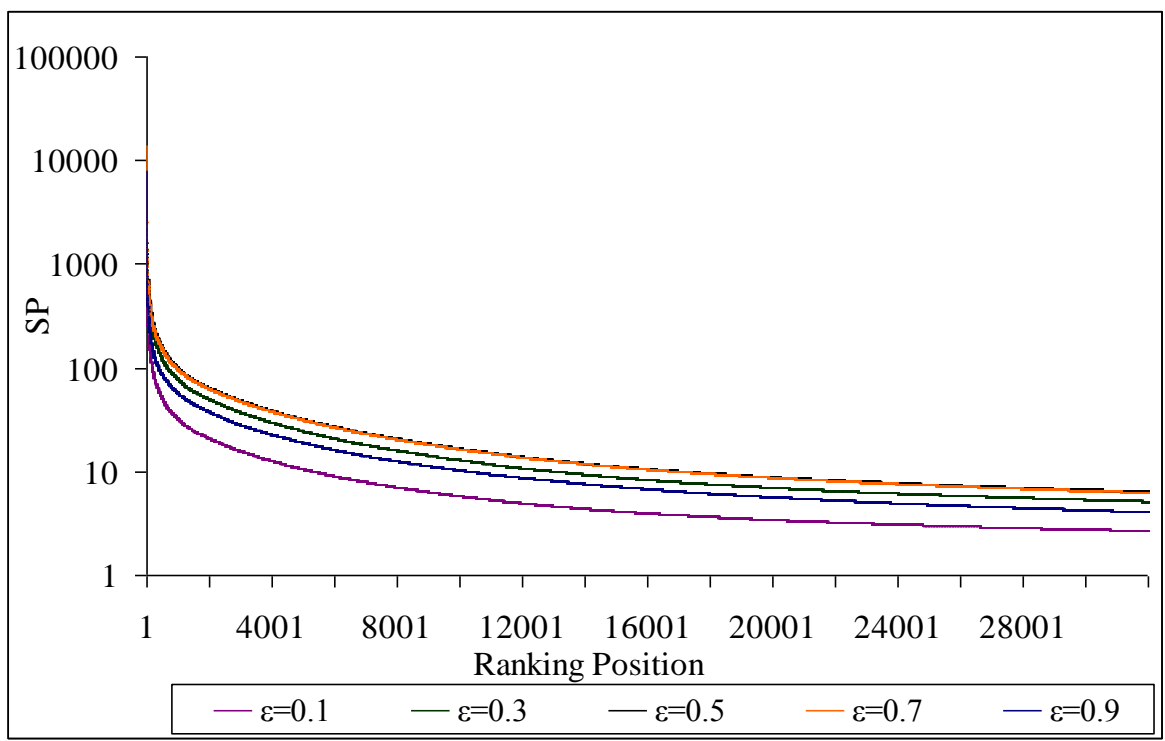

Figure 17 Social positions for the top 30,000 members within the BT dataset in relation to $\varepsilon$ where the commitment function was assessed based on the number of the phone calls

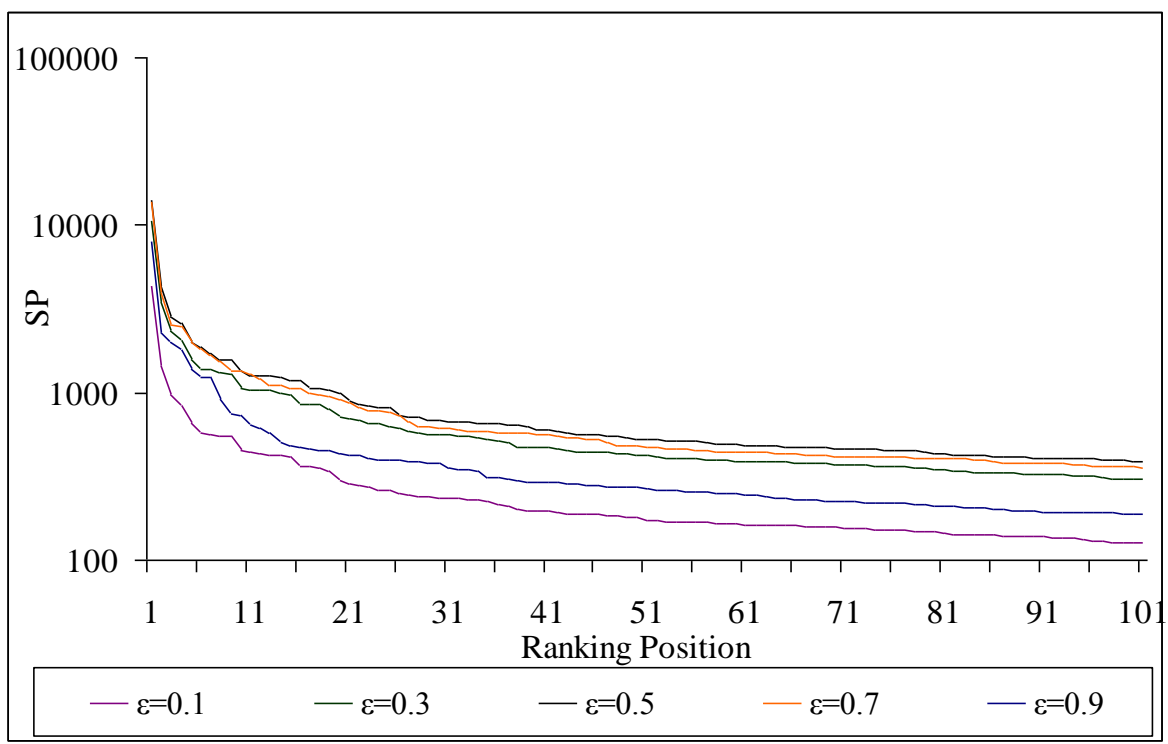

Figure 18 Social positions for the top 100 members within the BT dataset in relation to $\varepsilon$ where the commitment function was assessed based on the number of the phone calls

Moreover the social positions values for the top 100 users are presented in Figure 18

(where the commitment function was assessed based on the number of the phone calls) and Figure 20 (where the commitment function was assessed based on the duration of the phone calls). Figures 19 and 21 show also that the rankings of the social positions are similar for different $\varepsilon$ values. For more detailed analysis see section 6.3 Ranking Comparison.

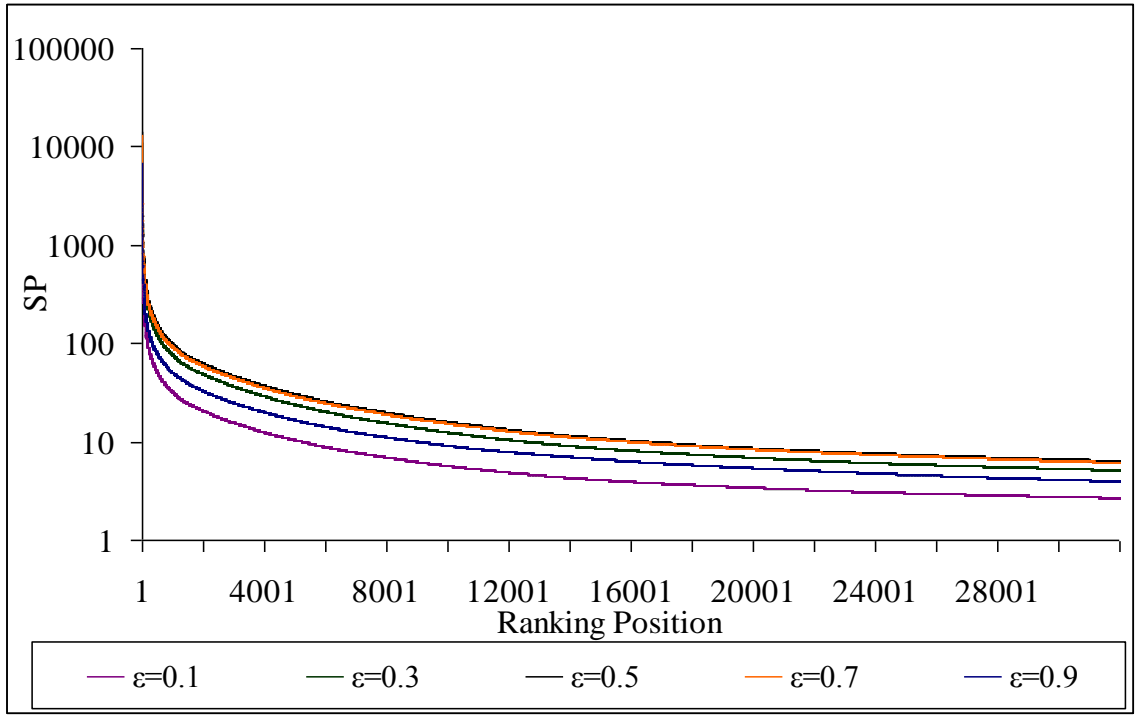

Figure 19 Social positions for the top 30,000 members within the BT dataset in relation to $\varepsilon$ where the commitment function was assessed based on the duration of the phone calls

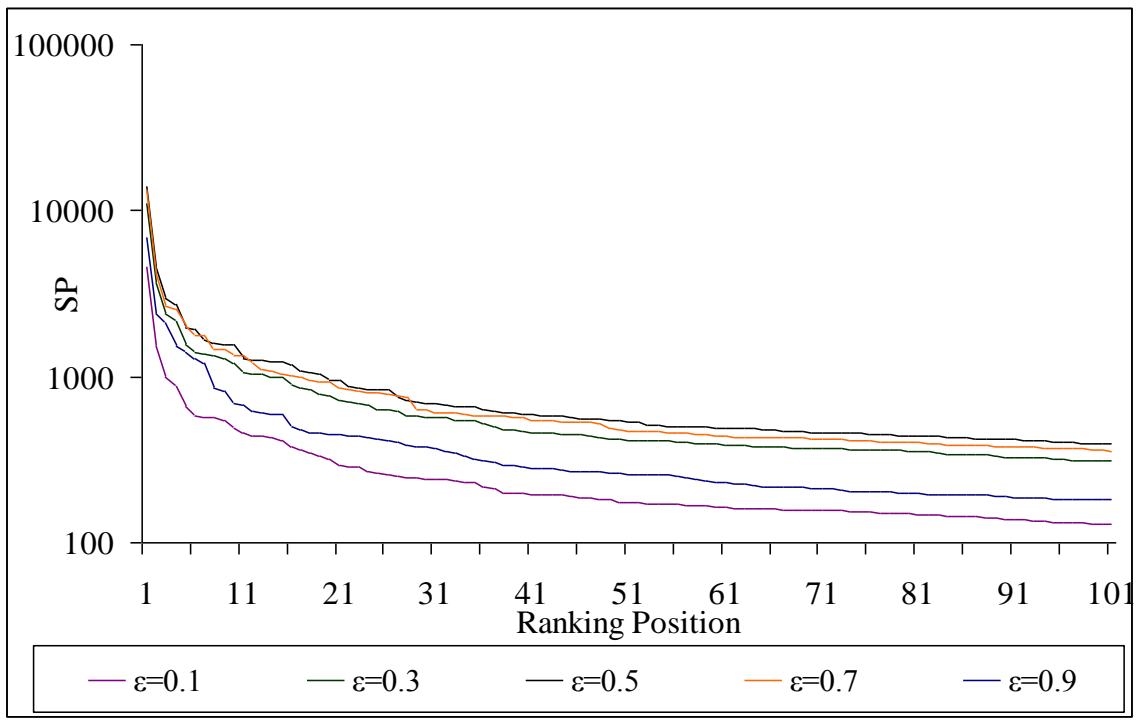

Figure 20 Social positions for the top 100 members within the BT dataset in relation to $\varepsilon$ where the commitment function was assessed based on the duration of the phone calls

## 6.3 Rankings Comparison

The values of the social position measure, calculated for the BT dataset where the commitment function was evaluated based on the phone calls and $\varepsilon$ equals 0.5, is utilized to create a ranking list of network members. The obtained values of social position are utilized to order network members, i.e. the individuals with the highest social position are placed at the top whereas the members with the lowest social position occupy the last position in the ranking. Two members with the same social position are assigned the same, higher position in the ranking and the following ranking position remains empty. The same procedure of ranking creation has been applied to rankings based on indegree centrality *IC* and outdegree centrality *OC*.

Once ranking lists have been created, a method to compare them is needed. For this purpose Kendall's coefficient of concordance was used to determine the similarity between two ranking lists. Let *X* and *Y* be any *n*-item rankings, then Kendall's coefficient of concordance $K(X,Y)$ can be evaluated from the following formula [35]:

$$K(X,Y) = \frac{1}{n(n-1)} \sum_{i=1}^{n} \sum_{j=1}^{n} \text{sgn}(x_j - x_i) \cdot \text{sgn}(y_j - y_i), \quad (23)$$

where:

$x_i$ and $y_i$ are the positions of the same *i*th item in ranking *X* and *Y*, respectively; they range from 1 to *n*; $\text{sgn}(x_j-x_i)$ is the sign of the difference $x_j-x_i$. It means that if item *j* follows item *i* in ranking *X*, then $\text{sgn}(x_j-x_i)=+1$; if they are at the same position $\text{sgn}(x_j-x_i)=0$; otherwise $\text{sgn}(x_j-x_i)=-1$.

When two rankings have the same items in every position, Kendall's coefficient for them is equal to +1. However, when two rankings have all the same items but they occur in reverse order, their Kendall's coefficient equals −1.

### 6.3.1 Kendall's Coefficient Between Different Iterations

The main aim of which was to compare the entire rankings created upon social positions measures after 1st, 2nd, 3rd, 4th, 5th, and 6th iteration of SPIN algorithm for $\varepsilon=0.5$ for BT dataset. For each possible pair of rankings, Kendall's coefficient was calculated and the results are presented in Table 13 and Figure 21. The rankings of the social position values after the 1st, 2nd, 3rd, 4th, 5th, and 6th iteration are very similar to each other, Kendall's coefficient is very high: $0.816<K(X,Y)<0.983$. Moreover, the most similar rankings are these obtained after 3rd, 4th, 5th, and 6th iterations.

|  | 1st iteration | 2nd iteration | 3rd iteration | 4th iteration | 5th iteration |
|---|---|---|---|---|---|
| 2nd iteration | 0.836 | | | | |
| 3rd iteration | 0.816 | 0.839 | | | |
| 4th iteration | 0.850 | 0.888 | 0.944 | | |
| 5th iteration | 0.842 | 0.874 | 0.960 | 0.983 | |
| 6th iteration | 0.847 | 0.880 | 0.952 | 0.991 | 0.992 |

Table 13 Kendall's coefficient for rankings of *SP* created after 1st, 2nd, 3rd, 4th, 5th, and 6th iteration of SPIN algorithm for $\varepsilon=0.5$

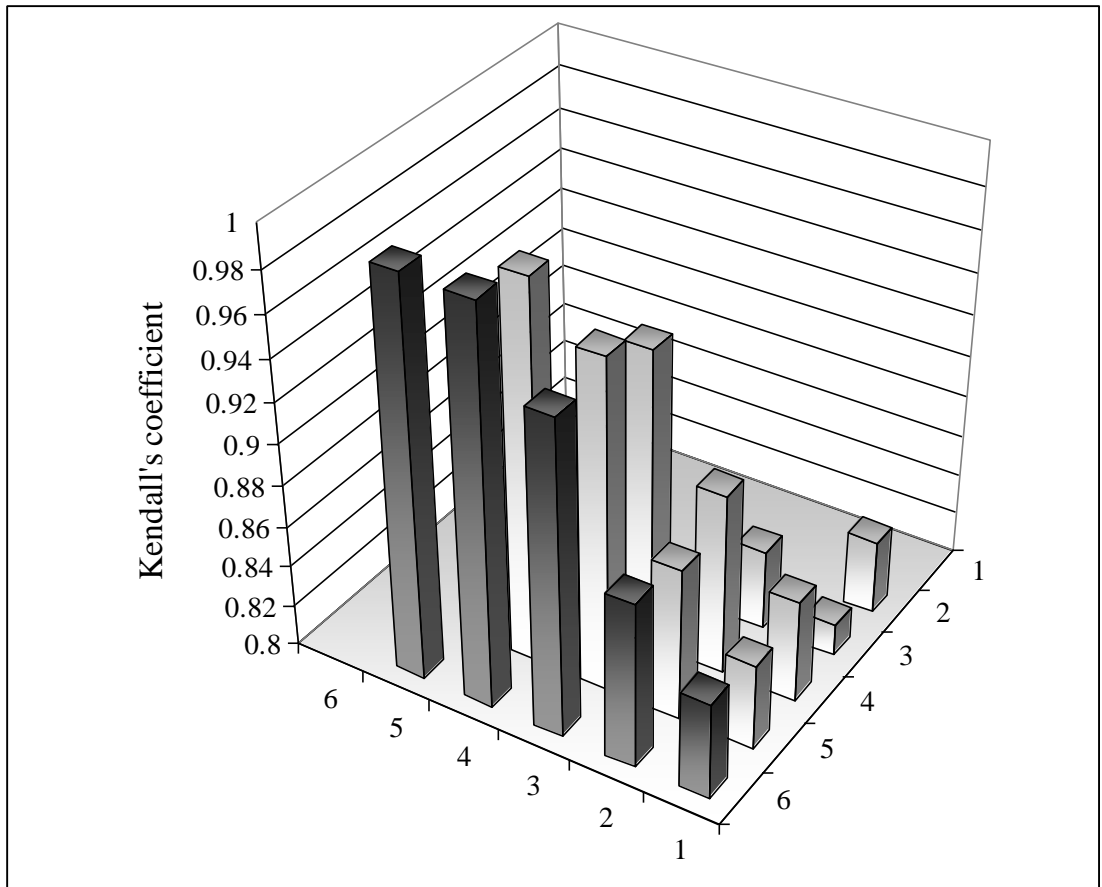

Figure 21 Kendall's coefficient for rankings of *SP* created after 1st, 2nd, 3rd, 4th, 5th, and 6th iteration of SPIN algorithm for $\varepsilon=0.5$

### 6.3.2 Social Position Rankings versus other Centrality Measures Rankings

The rankings of the social position values for different $\varepsilon$ are very similar to each other, Kendall's coefficient is very high: $0.786<K(X,Y)<0.911$; whereas the rankings based on indegree centrality (*IC*) and outdegree centrality (*OC*) differ a lot from all the others. Kendall's coefficient is relatively high between *IC* ranking and *SP* rankings and similar to all $\varepsilon$ values, i.e. $0.466<K(X,Y)<0.471$. On the other hand, the ranking of *OC* differs a lot both from *IC* ($K(X,Y)= -0.338$) and *SP* ($0.350<K(X,Y)<0.357$) (Table 14 and Figure 22).

|  | $\varepsilon=0.3$ | $\varepsilon=0.5$ | $\varepsilon=0.7$ | *IC* |
|---|---|---|---|---|
| $\varepsilon=0.5$ | 0.911 | | | |
| $\varepsilon=0.7$ | 0.786 | 0.867 | | |
| *IC* | 0.467 | 0.471 | 0.466 | |
| *OC* | -0.350 | -0.353 | -0.357 | -0.338 |

Table 14 Kendall's coefficient for rankings of *SP* for $\varepsilon=0.3$, $\varepsilon=0.5$, $\varepsilon=0.7$, Indegree (*IC*) and Outdegree (*OC*) Centrality

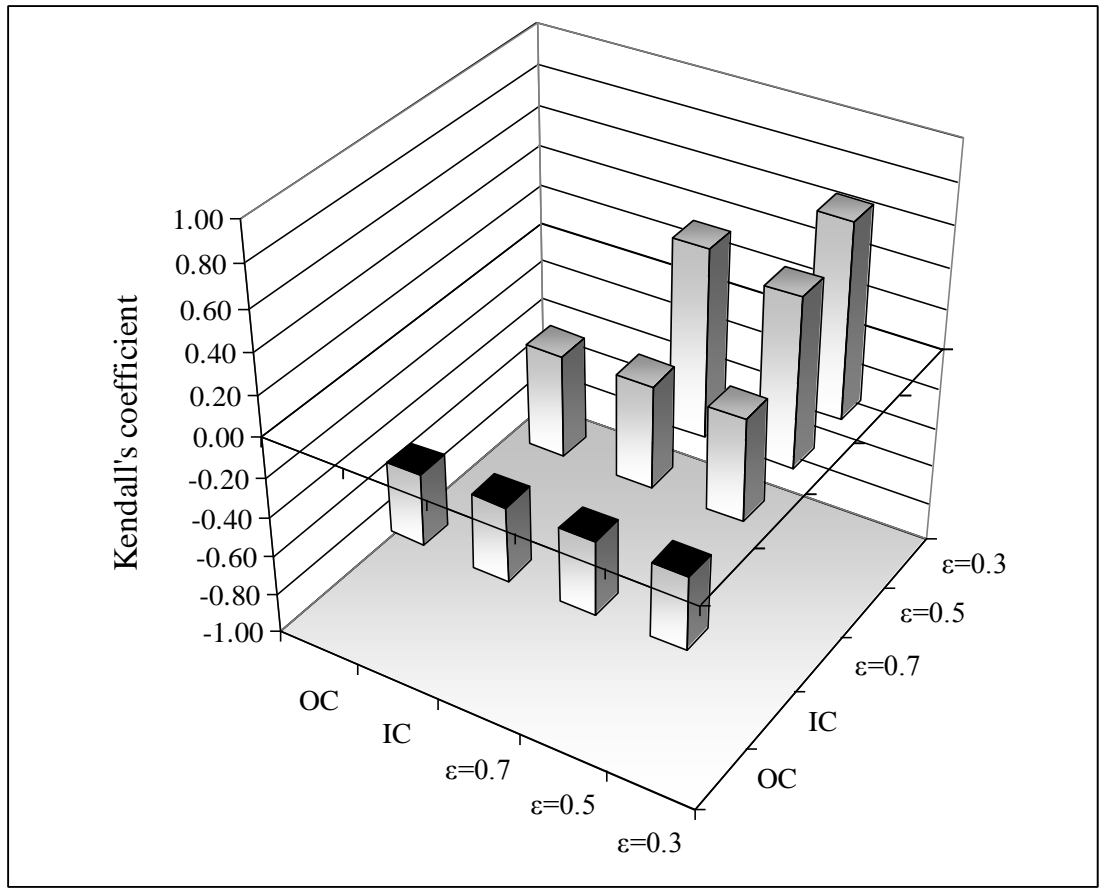

Figure 22 Kendall's coefficient for rankings of *SP* for $\varepsilon=0.3$, $\varepsilon=0.5$, $\varepsilon=0.7$, Indegree (*IC*) and Outdegree (*OC*) Centrality

## 6.4 Efficiency Tests

The main aim of the performed efficiency tests is to investigate, which of the three developed algorithms: SPIN$^{node}$, SPIN$^{edges}$ or SPIN$^{hybrid}$ is the most effective one. Efficiency is understood as the time it takes for an algorithm to calculate social position.

The efficiency tests are split into two main stages. First, the influence of $\varepsilon$ coefficient on processing time of different variants of SPIN algorithms is investigated. In the second phase, the tests are performed on the networks of different size, i.e. with different number of nodes and edges. These are random networks that were generated for the needs of the experiments.

### 6.4.1 Influence of ε on the Time Processing

This part of experiments was performed on BT dataset. The tests were done for the following $\varepsilon$ values: 0.1, 0.2, 0.3, 0.4, 0.5, 0.6, 0.7, 0.8, 0.9. For each of the enumerated $\varepsilon$ value the processing time of three developed algorithms (SPIN$^{node}$, SPIN$^{edges}$, and SPIN$^{hybrid}$) was calculated and the outcomes are presented in Table 15. The values in the table are the time (in seconds or in minutes) of one iteration for the given algorithm.

| $\varepsilon$ | 0.1 | 0.2 | 0.3 | 0.4 | 0.5 | 0.6 | 0.7 | 0.8 | 0.9 |
|---|---|---|---|---|---|---|---|---|---|
| **SPIN$^{node}$** | | | | | | | | | |
| time [s] | 338,88 | 338,58 | 338,69 | 338,74 | 338,76 | 338,85 | 338,68 | 338,40 | 339,23 |
| time [min] | 5,648.0 | 5,643.1 | 5,644.9 | 5,645.6 | 5,646.1 | 5,647.6 | 5,644.8 | 5,640.1 | 5,653.9 |
| **SPIN$^{edges}$** | | | | | | | | | |

| time [s]   | 2,646  | 2,740  | 2,790  | 2,769  | 2,800  | 2,749  | 2,757  | 2,722  | 2,703  |
|------------|--------|--------|--------|--------|--------|--------|--------|--------|--------|
| time [min] | 44.10  | 45.66  | 46.51  | 46.15  | 46.67  | 45.82  | 45.96  | 45.37  | 45.05  |
| **SPIN$^{hybrid}$** | | | | | | | | | |
| time [s]   | 6,764  | 6,663  | 6,651  | 6,739  | 6,779  | 6,775  | 6,664  | 6,746  | 6,679  |
| time [min] | 112.73 | 111.04 | 110.86 | 112.32 | 112.99 | 112.92 | 111.07 | 112.43 | 111.32 |

Table 15 Time processing of the SPIN algorithm depending on different values of $\varepsilon$ coefficient

It can be easily noticed that the processing time is the biggest for SPIN$^{node}$ version of the algorithm and the shortest for the SPIN$^{edges}$ one. SPIN$^{edges}$ is over 120 times faster than SPIN$^{node}$ algorithm and about 2.5 times faster than the SPIN$^{hybrid}$.

|              | Average time of one iteration [min] | Standard deviation [min] |
|--------------|-------------------------------------|--------------------------|
| SPIN$^{nodes}$  | 5646.04                             | 3.79                     |
| SPIN$^{edges}$  | 45.70                               | 0.79                     |
| SPIN$^{hybrid}$ | 111.96                              | 0.88                     |

Table 16 Average time processing of one iteration and its standard deviation for different SPIN algorithms

When the $\varepsilon$ coefficient is taken into consideration then the average processing time of one iteration for SPIN$^{nodes}$ is over 5000 minutes, for SPIN$^{edges}$ is aroung 41 minutes and for SPIN$^{hybrid}$ equals 100 minutes (Table 16, Figure 23). The analysis of standard deviation of these values enables to assess how the $\varepsilon$ coefficient influence the time processing of SPIN algorithms (Table 16, Figure 24). The smallest standard deviation is in the case of SPIN$^{edges}$ algorithm and it equals 0.79 [min] whereas the biggest one is for SPIN$^{nodes}$ (3.79 [min]) and this is intuitive because the average time of one iteration is also the biggest. These standard deviations are small in comparison to average processing time of one iteration for different $\varepsilon$ values so it can be assumed that the value of $\varepsilon$ coefficient does not influence the time processing of the algorithms.

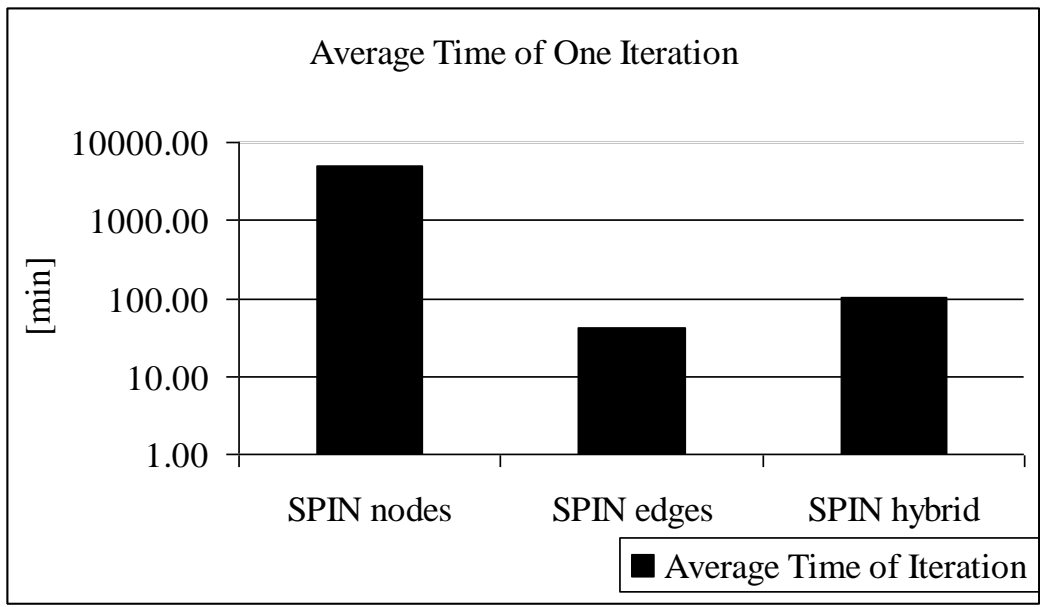

Figure 23 Average time procession of one iteration for different variants of SPIN algorithm

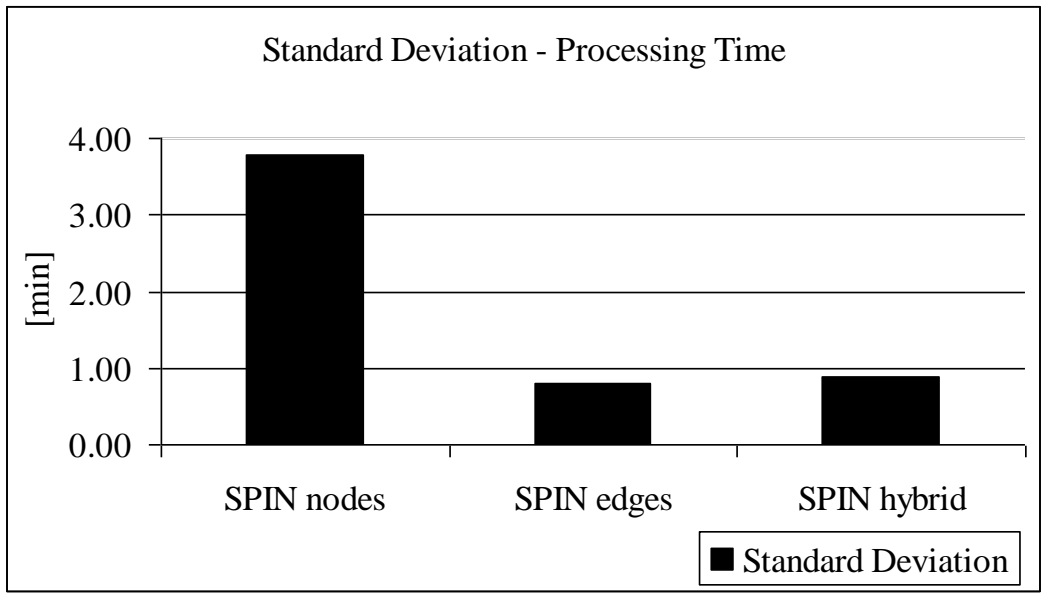
Figure 24 Standard deviation of time procession of one iteration for different variants of SPIN algorithm

### 6.4.2 Influence of Network Size on Processing Time

The next stage of the efficiency test were performed on randomly generated social networks. In order to do that the algorithm that serves to generate the random networks were developed and 25 different directed networks were generated (Table 17). All further experiments in this section were performed for $\varepsilon=0.8$ as it was shown in section 6.4.1 that the value of $\varepsilon$ value does not influence the processing time.

| ID | Number of Nodes | Number of Edges |
|---|---|---|
| 1 | 1,000 | 1,000 |
| 2 |  | 5,000 |
| 3 |  | 10,000 |
| 4 |  | 50,000 |
| 5 |  | 100,000 |
| 6 | 5,000 | 1,000 |
| 7 |  | 5,000 |
| 8 |  | 10,000 |
| 9 |  | 50,000 |
| 10 |  | 100,000 |
| 11 | 10,000 | 1,000 |
| 12 |  | 5,000 |
| 13 |  | 10,000 |
| 14 |  | 50,000 |
| 15 |  | 100,000 |
| 16 | 50,000 | 1,000 |
| 17 |  | 5,000 |
| 18 |  | 10,000 |
| 19 |  | 50,000 |
| 20 |  | 100,000 |
| 21 | 100,000 | 1,000 |

| ID | Number of Nodes | Number of Edges |
|----|-----------------|-----------------|
| 22 |                 | 5,000           |
| 23 |                 | 10,000          |
| 24 |                 | 50,000          |
| 25 |                 | 100,000         |

Table 17 The random networks generated for the needs of the efficiency tests

First, the tests were performed for SPIN$^{nodes}$ algorithm (Table 18 and Figure 25). The processing time for the biggest network (100,000 nodes and the same number of edges) was approximately 1950 times longer than the processing time for the smallest network (1,000 nodes and 1,000 edges). It reveals that the network size has great influence on time processing. The bigger network, the processing time increases a lot.

| Edges \ Nodes | 1,000 | 5,000 | 10,000 | 50,000 | 100,000 | Unit |
|---|---|---|---|---|---|---|
| 1,000 | 15,536.67 | 27,781.33 | 39,963.67 | 138,765.67 | 255,041.67 | [ms] |
|  | 15.54 | 27.78 | 39.96 | 138.77 | 255.04 | [s] |
|  | 0.26 | 0.46 | 0.67 | 2.31 | 4.25 | [min] |
| 5,000 | 265,500.00 | 326,593.50 | 392,547.00 | 861,687.50 | 1,444,969.00 | [ms] |
|  | 265.50 | 326.59 | 392.55 | 861.69 | 1444.97 | [s] |
|  | 4.43 | 5.44 | 6.54 | 14.36 | 24.08 | [min] |
| 10,000 | 976,547.00 | 1,144,601.50 | 1,189,015.50 | 2,160,859.00 | 3,367,281.50 | [ms] |
|  | 976.55 | 1,144.60 | 1,189.02 | 2,160.86 | 3,367.28 | [s] |
|  | 16.28 | 19.08 | 19.82 | 36.01 | 56.12 | [min] |
| 50,000 | 10,538,828.00 | 10,937,891.00 | 11,241,656.00 | 14,701,078.00 | 17,517,500.00 | [ms] |
|  | 10,538.83 | 10,937.89 | 11,241.66 | 14,701.08 | 17,517.50 | [s] |
|  | 175.65 | 182.30 | 187.36 | 245.02 | 291.96 | [min] |
| 100,000 | 22,141,313.00 | 22,185,781.00 | 23,360,890.00 | 26,917,687.00 | 30,304,938.00 | [ms] |
|  | 22,141.31 | 22,185.78 | 23,360.89 | 26917.69 | 30,304.94 | [s] |
|  | 369.02 | 369.76 | 389.35 | 448.63 | 505.08 | [min] |

Table 18 Time processing of the SPIN$^{nodes}$ depending on different size of network

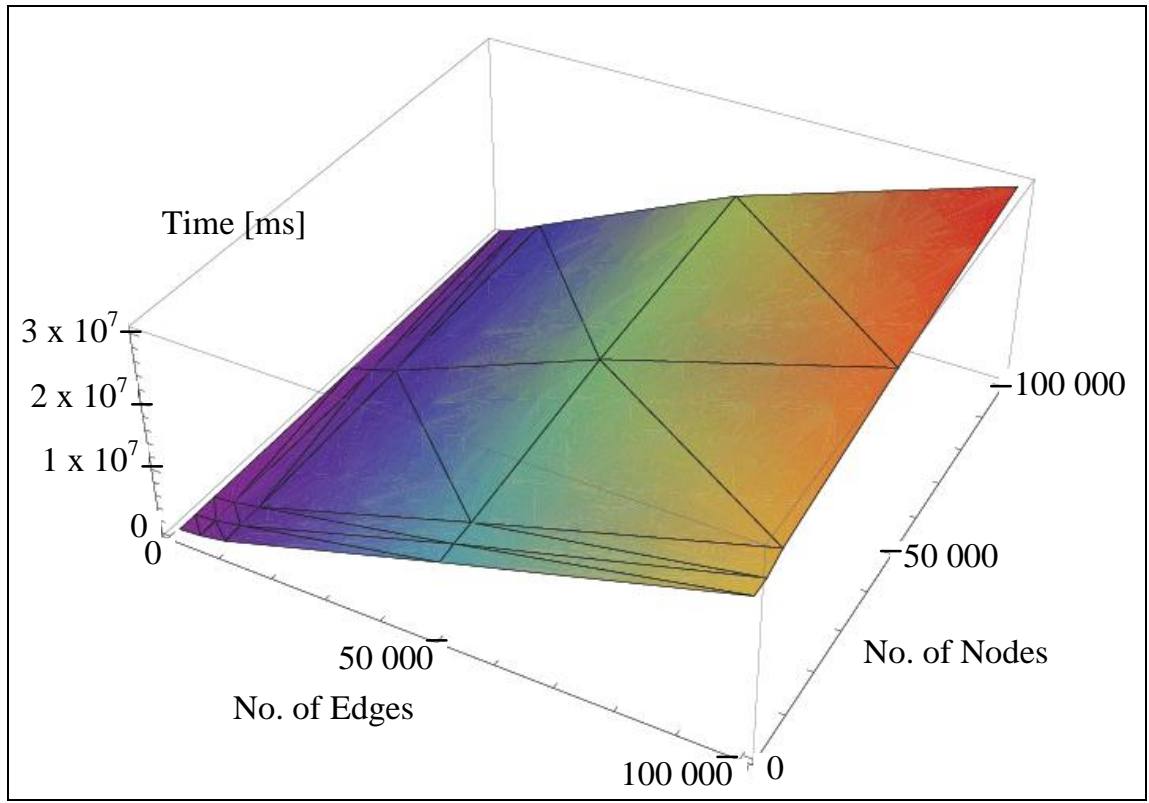

Figure 25 Time processing of the SPIN$^{node}$ depending on different size of network

The next tests were performed for SPIN$^{edges}$ algorithm (Table 19 and Figure 26). The processing time for the biggest network (100,000 nodes and the same number of edges) was approximately 84 times longer than the processing time for the smallest network (1,000 nodes and 1,000 edges). It shows that the influence of the network size on processing time is smaller than in the case of the SPIN$^{nodes}$ algorithm.

| Edges \ Nodes | 1000 | 5000 | 10000 | 50000 | 100000 | Unit |
|---|---|---|---|---|---|---|
| 1,000 | 447.83 | 734.50 | 1,052.00 | 3,552.00 | 6,119.83 | [ms] |
|  | 0.45 | 0.73 | 1.05 | 3.55 | 6.12 | [s] |
|  | 0.01 | 0.01 | 0.02 | 0.06 | 0.10 | [min] |
| 5,000 | 1,721.33 | 2,104.17 | 2,421.83 | 5,041.67 | 8,132.83 | [ms] |
|  | 1.72 | 2.10 | 2.42 | 5.04 | 8.13 | [s] |
|  | 0.03 | 0.04 | 0.04 | 0.08 | 0.14 | [min] |
| 10,000 | 3,455.67 | 3,812.67 | 3,960.83 | 6,554.50 | 10,093.67 | [ms] |
|  | 3.46 | 3.81 | 3.96 | 6.55 | 10.09 | [s] |
|  | 0.06 | 0.06 | 0.07 | 0.11 | 0.17 | [min] |
| 50,000 | 16,565.33 | 16,195.17 | 16,388.00 | 18,953.17 | 22,786.33 | [ms] |
|  | 16.57 | 16.20 | 16.39 | 18.95 | 22.79 | [s] |
|  | 0.28 | 0.27 | 0.27 | 0.32 | 0.38 | [min] |
| 100,000 | 31,971.50 | 31,940.33 | 33,033.83 | 35,924.67 | 37,924.67 | [ms] |
|  | 31.97 | 31.94 | 33.03 | 35.92 | 37.92 | [s] |
|  | 0.53 | 0.53 | 0.55 | 0.60 | 0.63 | [min] |

Table 19 Time processing of the SPIN$^{edges}$ depending on different size of network

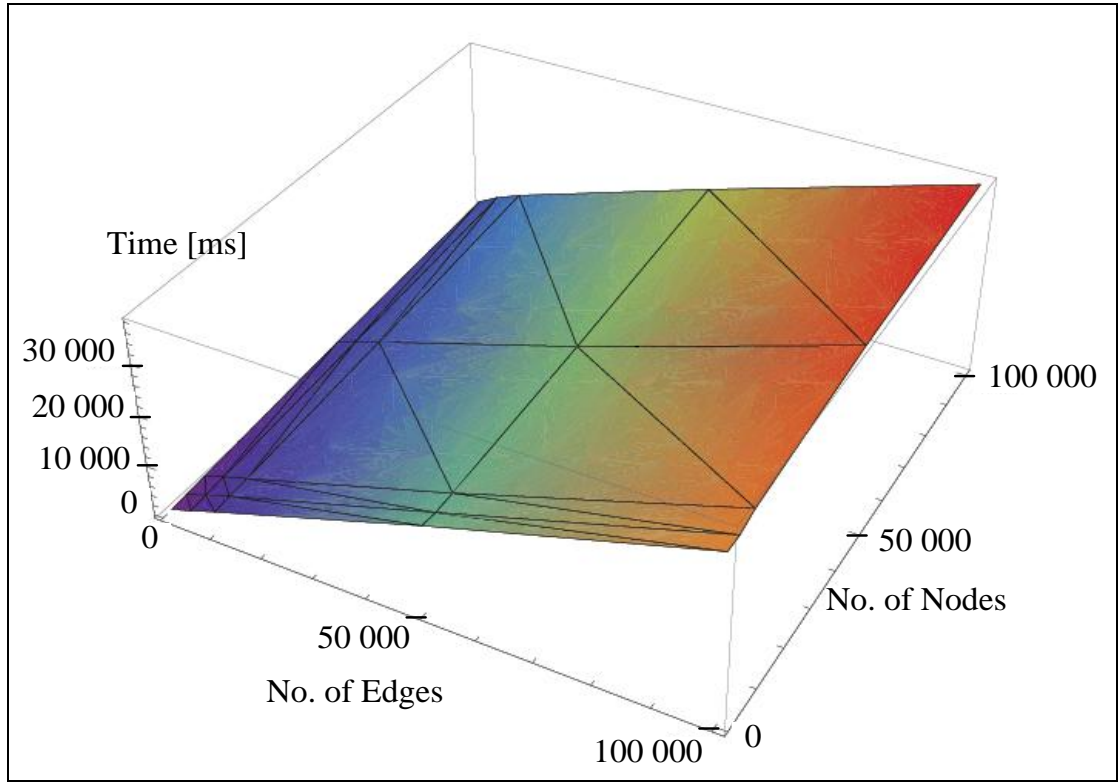

Figure 26 Time processing of the SPIN$^{edges}$ depending on different size of network

The last tests were performed for SPIN$^{hybrid}$ algorithm (Table 20 and Figure 27). The processing time for the biggest network (100,000 nodes and the same number of edges) was approximately 88 times longer than the processing time for the smallest network (1,000 nodes and 1,000 edges). Similarly to SPIN$^{edges}$ it shows that the influence of the network size on processing time is smaller than in the case of the SPIN$^{nodes}$ algorithm. Moreover the influence of network size is similar when SPIN$^{edges}$ and SPIN$^{nodes}$ algorithms are considered.

| Edges \ Nodes | 1,000 | 5,000 | 10,000 | 50,000 | 100,000 | Unit |
|---|---|---|---|---|---|---|
| 1,000 | 906.20 | 1,159.40 | 1,425.00 | 3,734.60 | 6,790.80 | [ms] |
|  | 0.91 | 1.16 | 1.43 | 3.73 | 6.79 | [s] |
|  | 0.02 | 0.02 | 0.02 | 0.06 | 0.11 | [min] |
| 5,000 | 3,756.20 | 4,025.00 | 4,375.00 | 7,306.20 | 9,843.80 | [ms] |
|  | 3.76 | 4.03 | 4.38 | 7.31 | 9.84 | [s] |
|  | 0.06 | 0.07 | 0.07 | 0.12 | 0.16 | [min] |
| 10,000 | 7,587.40 | 7,806.20 | 7,897.00 | 10,425.00 | 13,934.40 | [ms] |
|  | 7.59 | 7.81 | 7.90 | 10.43 | 13.93 | [s] |
|  | 0.13 | 0.13 | 0.13 | 0.17 | 0.23 | [min] |
| 50,000 | 35,765.60 | 35,506.20 | 35,668.80 | 38,896.80 | 43,837.40 | [ms] |
|  | 35.77 | 35.51 | 35.67 | 38.90 | 43.84 | [s] |
|  | 0.60 | 0.59 | 0.59 | 0.65 | 0.73 | [min] |
| 100,000 | 69,444.00 | 70,568.80 | 71,503.20 | 76,869.00 | 80,006.20 | [ms] |
|  | 69.44 | 70.57 | 71.50 | 76.87 | 80.01 | [s] |
|  | 1.16 | 1.18 | 1.19 | 1.28 | 1.33 | [min] |

Table 20 Time processing of the SPIN$^{hybrid}$ depending on different size of network

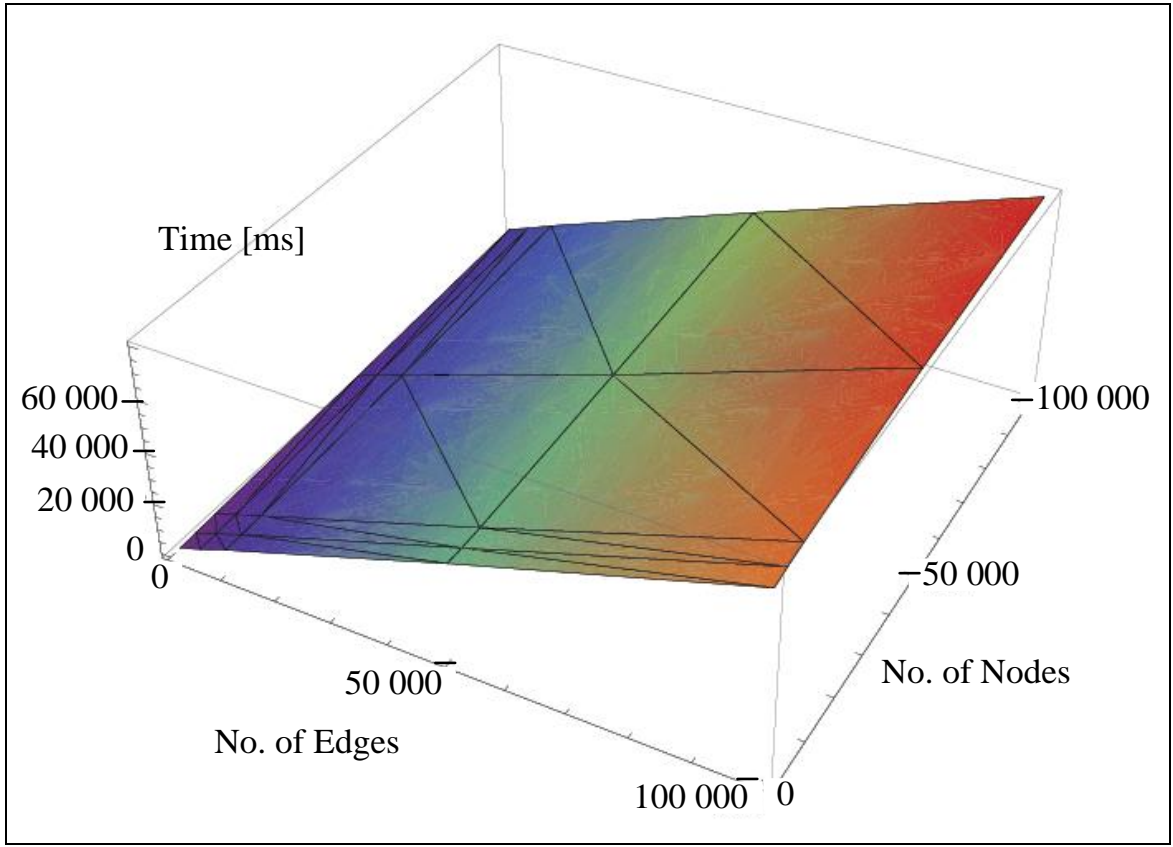

Figure 27 Time processing of the SPIN$^{hybrid}$ depending on different size of network

The comparison of different variants of SPIN algorithm reveals that the fastest one is always SPIN$^{edges}$. Consider for example processing time for networks where the number of edges is constant and equals 50,000 whereas the number of nodes changes as shown in Table 21. Note that in case of the SPIN$^{edges}$ and SPIN$^{hybrid}$ algorithms the processing times do not differ a lot among 1,000, 5,000 and 10,000 nodes and it oscillates around 16 [s] for SPIN$^{edges}$ and 35 [s] for SPIN$^{hybrid}$.

| No. of Nodes | SPIN$^{nodes}$ [s] | SPIN$^{edges}$ [s] | SPIN$^{hybrid}$ [s] |
|---|---|---|---|
| 1,000 | 10,539 | 16.56 | 35.77 |
| 5,000 | 10,938 | 16.19 | 35.51 |
| 10,000 | 11,242 | 16.39 | 35.67 |
| 50,000 | 14,701 | 18.95 | 38.90 |
| 100,000 | 17,518 | 22.79 | 43.84 |

Table 21 Processing time in relation to the number of nodes in the network for fixed number of edges (50,000)

The SPIN$^{edges}$ is 636.2 times faster than SPIN$^{nodes}$ for 1,000 nodes and 768.77 times faster for 100,000 nodes. Simultaneously, SPIN$^{edges}$ algorithm is approximately two times faster than SPIN$^{hybrid}$ algorithm for all types of investigated random networks where number of edges equals 50,000 (Table 22).

| No. of Nodes | $\dfrac{t_{SPIN^{nodes}}}{t_{SPIN^{edges}}}$ | $\dfrac{t_{SPIN^{hybrid}}}{t_{SPIN^{edges}}}$ |
|---|---|---|
| 1,000 | 636.20 | 2.16 |
| 5,000 | 675.38 | 2.19 |
| 10,000 | 685.97 | 2.18 |
| 50,000 | 775.65 | 2.05 |
| 100,000 | 768.77 | 1.92 |

Table 22 The relation of processing times of $SPIN^{edges}$ to other SPIN algorithms for fixed number of edges (50,000)

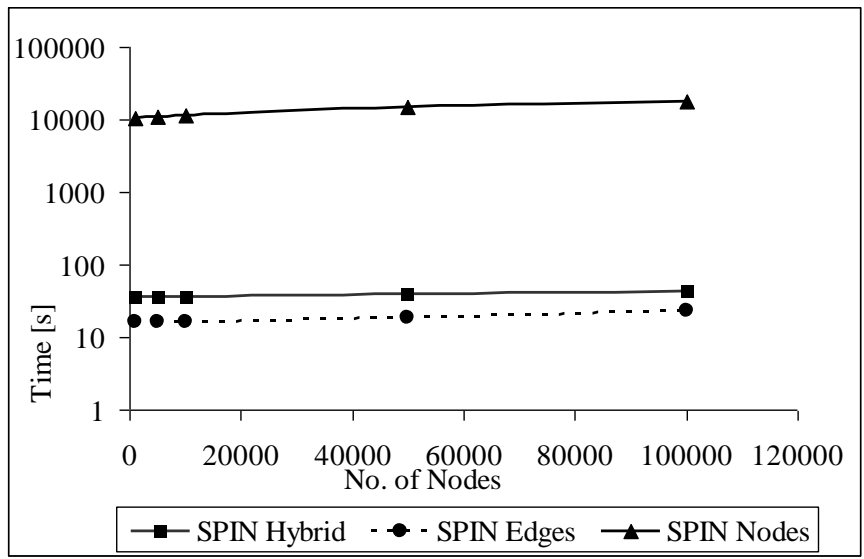

Figure 28 Time processing depending on the number of nodes in the network for fixed number of edges (50,000)

The Figure 28 shows that the processing time is a monotonically and increasing function of the number of nodes in the network, i.e. the greater number of nodes, the greater processing time. However only in case of the $SPIN^{edges}$ algorithm the processing time is additionally linear function of the number of nodes in the network. Moreover, the tangent of slope angle is very close to zero and it means that the values of the function increase very slow. In other words there are almost constant (Table 23).

| No. of Nodes | $SPIN^{nodes}$ | $SPIN^{edges}$ | $SPIN^{hybrid}$ |
|---|---|---|---|
| 1,000 | 10.5388 | 0.0016 | 2.1591 |
| 5,000 | 2.1876 | 0.0015 | 2.1924 |
| 10,000 | 1.1242 | 0.0015 | 2.1765 |
| 50,000 | 0.2940 | 0.0013 | 2.0523 |
| 100,000 | 0.1752 | 0.0013 | 1.9238 |

Table 23 The ratio of processing time and number of nodes for different SPIN algorithms for constant number of edges (50,000)

On the other hand let's consider the processing time for networks where the number of nodes is constant and equals 50,000 whereas the number of edges changes as shown in Table

24. Note that, in contrary to networks when the number of edges is constant, in case of the SPIN$^{edges}$ and SPIN$^{hybrid}$ algorithms the processing times differ a lot among 1,000, 5,000, 10,000, 50,000 and 100,000 edges. It changes from 3.55 [s] for 1,000 edges to 35.92 [s] for 100,000edges for SPIN$^{edges}$. Additionally, it changes from 3.73 [s] for 1,000 edges to 76.87 [s] for 100,000edges for SPIN$^{hybrids}$.

| No. of Edges | SPIN$^{nodes}$ [s] | SPIN$^{edges}$ [s] | SPIN$^{hybrid}$ [s] |
|---|---|---|---|
| 1,000 | 138.77 | 3.55 | 3.73 |
| 5,000 | 861.69 | 5.04 | 7.31 |
| 10,000 | 2,160.86 | 6.55 | 10.43 |
| 50,000 | 14,701.08 | 18.95 | 38.90 |
| 100,000 | 26,917.69 | 35.92 | 76.87 |

Table 24 Time processing depending on the number of nodes in the network for fixed number of nodes (50,000)

The SPIN$^{edges}$ is 39.07 times faster than SPIN$^{nodes}$ for 1,000 nodes and 749.28 times faster for 100,000 nodes. Simultaneously, SPIN$^{edges}$ algorithm is one time faster than SPIN$^{hybrid}$ for 1,000 nodes and 2 times faster for 100,000 nodes. (Table 25).

| No. of Nodes | $\dfrac{t_{SPIN^{nodes}}}{t_{SPIN^{edges}}}$ | $\dfrac{t_{SPIN^{hybrid}}}{t_{SPIN^{edges}}}$ |
|---|---|---|
| 1,000 | 39.07 | 1.05 |
| 5,000 | 170.91 | 1.45 |
| 10,000 | 329.68 | 1.59 |
| 50,000 | 775.65 | 2.05 |
| 100,000 | 749.28 | 2.14 |

Table 25 The relation of processing times of SPIN$^{edges}$ to other SPIN algorithms for fixed number of nodes (50,000)

The Figure 29 shows that the processing time is a monotonically and increasing function of the number of nodes in the network, i.e. the greater number of nodes, the greater processing time. However none of them can be seen as the linear function of the number of nodes in the network (Table 26).

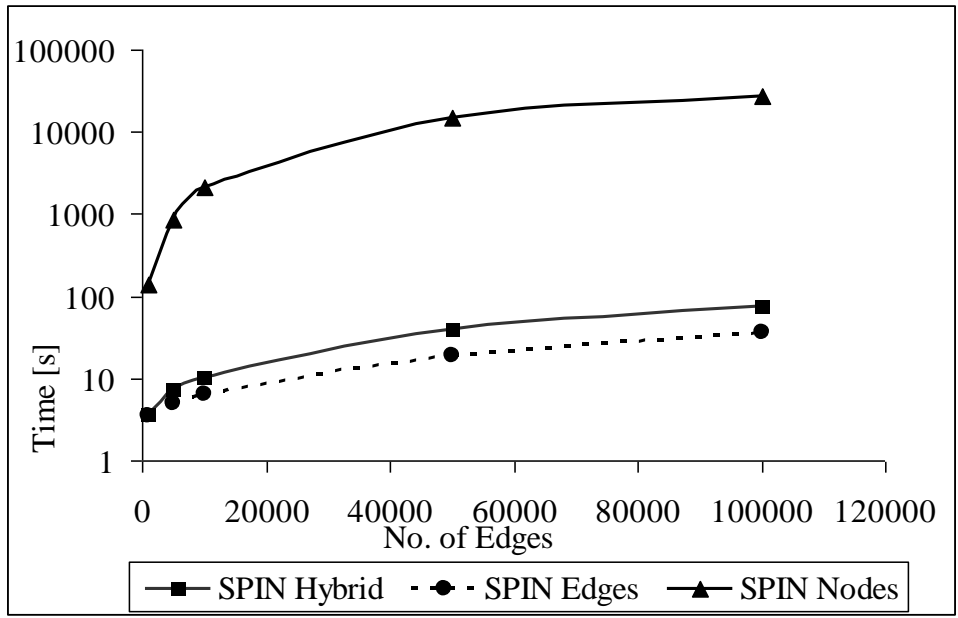

Figure 29 Time processing depending on the number of edges in the network for fixed number of nodes (50,000)

| No. of Edges | SPIN$^{nodes}$ | SPIN$^{edges}$ | SPIN$^{hybrid}$ |
|---|---|---|---|
| 1,000 | 0.1388 | 0.0256 | 1.0514 |
| 5,000 | 0.1723 | 0.0059 | 1.4492 |
| 10,000 | 0.2161 | 0.0030 | 1.5905 |
| 50,000 | 0.2940 | 0.0013 | 2.0523 |
| 100,000 | 0.2692 | 0.0013 | 2.1397 |

Table 26 The ratio of processing time and number of edges for different SPIN algorithms for constant number of nodes (50,000)

## 6.5 Social Position versus Other Centrality Indices

The goal of the last part of experiments is to compare the proposed social position measure to other indices that serve to assess the position of the user within the social network. The measures that are taken into consideration are indegree centrality (*IC*) and outdegree centrality (*OC*) (see section 2.5 for detailed description). First, the processing time of the used algorithms are analyzed. After that, the number of duplicates, i.e. the number of users that possess the same centrality value, is investigated. Moreover, the rankings for *SP*, *IC*, and *OC* were compared in section 6.3.2.

### 6.5.1 Efficiency Comparison

The efficiency tests were performed for BT dataset in which the number of nodes is 4,455,802 and number of edges equals 17,063,810 and the outcomes are presented in Table 23. In order to compare the SPIN algorithm with indegree and outdegree centrality algorithms, the appropriate versions of *IC* and *OC* algorithms were developed and in consequence the SPIN$^{edges}$ is compared with OC$^{edges}$ and IC$^{edges}$ and SPIN$^{hybrid}$ is compared with OC$^{hybrid}$ and IC$^{hybrid}$.

|  | SPIN$^{edges}$ | SPIN$^{hybrid}$ | OC$^{edges}$ | OC$^{hybrid}$ | IC$^{edges}$ | IC$^{hybrid}$ |
|---|---|---|---|---|---|---|
| time [s] | 2,722.35 | 6,745.54 | 2,425.81 | 3,265.77 | 2,520.90 | 3,804.93 |
| time [min] | 45.37 | 112.43 | 40.43 | 54.43 | 42.02 | 63.42 |

Table 27 Average processing time of the SPIN algorithm in comparison with outdegree (OC) and indegree (IC) centrality

The outcomes for the 'hybrid' versions of the algorithms shows that SPIN processing time is two times longer than the IC and OC. More detailed analysis of 'edges' versions of centrality algorithms reveals that all of them are comparable with respect to processing time (Table 23). It means that SPIN$^{edges}$ last 45.37 [min] whereas OC$^{edges}$ last 40.43 [min] and IC$^{edges}$ – 42.02 [min]. The fact that should be emphasized is that indegree and outdegree centralities take into consideration only the first level neighbors whereas the social position of user depends on the positions of all users within the network. This causes that the social position measure is much more diverse than other centrality indices (see section 6.5.2). Moreover indegree and outdegree are the simplest ones to calculate from centrality indicates, closeness and betweenness are much more difficult to compute.

### 6.5.2 Distribution and Number of Duplicates

Social position measure appears to be more diverse in comparison to other investigated in this thesis measures. It can be visible especially while analyzing number of users than possess the same centrality value (Figure 30 and 31, Table 28 and 29). These test were also performed for BT dataset .

The percentage of duplicates within the social positions measures, where commitment was evaluated based on the number of phone calls, is lower for each value of $\varepsilon$ in comparison with indegree and outdegree centrality. The social position measures generate 60% of duplicates after six iterations and for three values of $\varepsilon$ coefficient: 0.3, 0.5, 0.7 . The number of duplicates diminishes when more number of iterations is performed [31].

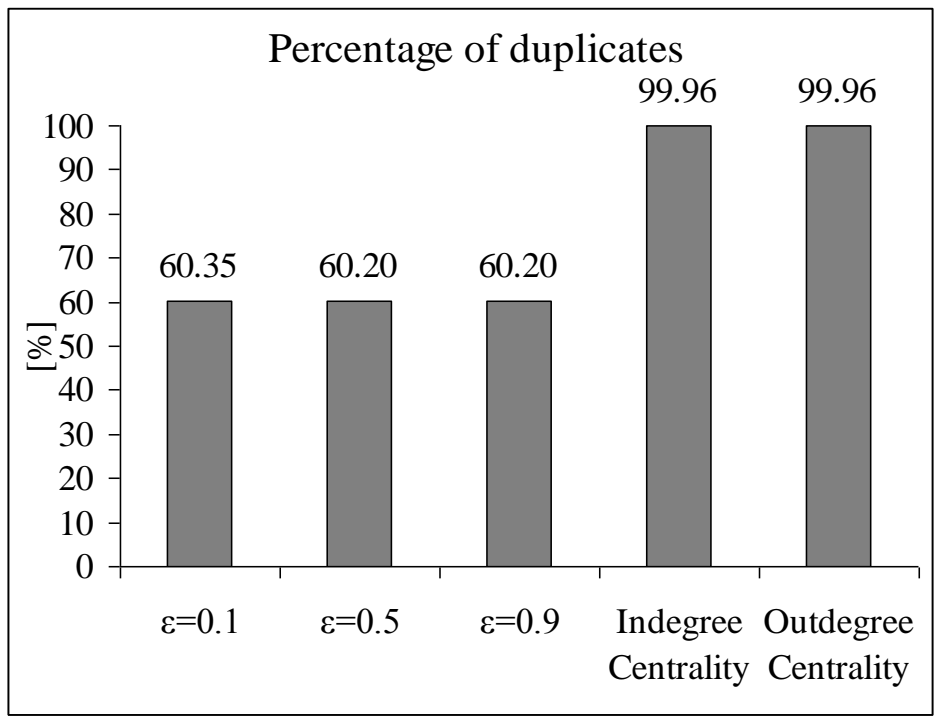

Figure 30 Percentage of duplicates within the set of node measures, separately for social position (where commitment was evaluated based on the number of phone calls) with different values of $\varepsilon$, indegree centrality, and outdegree centrality

In the analyzed network there are 4,455,802 nodes and it means that almost 2 mln users have distinct values of social positions whereas only 1,640 users obtained different indegree centrality and 1,662 different outdegree centrality. For that reason, the percentage of duplicates exceeds 99% for degree measures whereas it is around 60% for social position measure where commitment was evaluated based on the number of phone calls.

|  | $\varepsilon=0.1$ | $\varepsilon=0.5$ | $\varepsilon=0.9$ | *IC* | *OC* |
|---|---|---|---|---|---|
| Number of duplicates | 2,689,108 | 2,682,477 | 2,682,287 | 4,454,162 | 4,454,140 |
| Percentage of duplicates | 60.35 | 60.20 | 60.20 | 99.96 | 99.96 |

Table 28 Number and percentage of duplicates within the set of node measures, separately for social position (where commitment was evaluated based on the number of phone calls) with different values of $\varepsilon$, indegree centrality (*IC*), and outdegree centrality (*OC*)

The percentage of duplicates within the social positions measures, where commitment was evaluated based on the duration of phone calls, is lower for each value of $\varepsilon$ in comparison with indegree and outdegree centrality. The social position measures generate 50% of duplicates after six iterations and for three values of $\varepsilon$ coefficient: 0.3, 0.5, 0.7. Moreover, the research reveals that there are 10% less duplicates when the commitment function, used in the process of social position assessment, was evaluated based on the duration of phone calls than based on the number of phone calls.

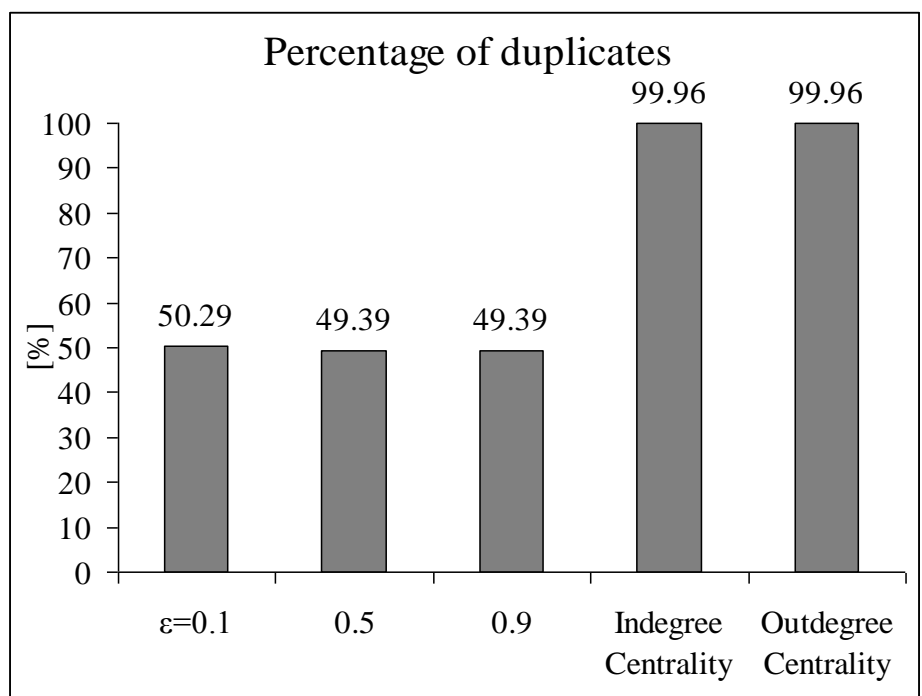

Figure 31 Percentage of duplicates within the set of node measures, separately for social position (where commitment was evaluated based on the duration of phone calls) with different values of $\varepsilon$, indegree centrality, and outdegree centrality

|  | $\varepsilon=0.1$ | $\varepsilon=0.5$ | $\varepsilon=0.9$ | *IC* | *OC* |
|---|---|---|---|---|---|
| Number of duplicates | 2,240,609 | 2,200,518 | 2,200,624 | 4,454,162 | 4,454,140 |
| Percentage of duplicates | 50.29 | 49.39 | 49.39 | 99.96 | 99.96 |

Table 29 Number and percentage of duplicates within the set of node measures, separately for social position (where commitment was evaluated based on the duration of phone calls) with different values of $\varepsilon$, indegree centrality (*IC*), and outdegree centrality (*OC*)

# 7 CONCLUSION

This master thesis presents the way of extracting key users within the customers of telecommunication services. The data were obtained from the Big European Telecomunication Provider plc company and based on them the social network of customers and their connections were created.

In order to answer research question no. 1 the social network had to be extracted from the pure telecommunication data. The appropriate and described in this work process (see section 5 Telecommunication Data) of data preparation was applied. First, the data has to be investigated in order to understand them and be able to define which users cannot be part the of social network i.e. all users which may not be people or social structures made by people so don't meet social network requirements e.g. machines, which call to people in order to play commercial or politician speech before election. The next very important stage is to cut down the number of users to the subset which is interesting from the research point of view. For example if the company needs to analyze the behaviour of their own users they should limit the data to their subscribers. On the other hand if company wants to make advertisement campaign and find potentially new clients they should focus only on users which aren't their subscribers. So different subsets can be considered and they depend on the research purpose. Wise limiting users number is very important because smaller number of people results in smaller network and in consequence the process of calculations is much faster (see section 6.4.2 Influence of Network Size on Processing Time). Of course possessing appropriate resources and calculation power may make subset extraction may not be necessary. In order to answer research question no. 2 the data, which can be used during commitment functions creation process. The data were: number of phone calls that users made and the duration of these phone calls. In consequence, two different commitment functions were proposed and investigated, i.e. one based on the number of phone calls and the other one based on the duration of these calls (for more details see sections 5.2 Data Pre-processing and 3.1 Commitment Function Evaluation). The data that was available but not used in this thesis were date of the phone call and time of the phone call. Nevertheless, the author is aware of the fact that this information can be useful when analyzing the dynamic of the social position measure and it will be developed in the future. Moreover, the complex commitment functions can be create based on date and time e.g. number of calls made in the particular time of the day i.e. in the morning, in the afternoon etc. to find when the users are most active in order to propose them additional free minutes or decrease cost of call to persuade user to stay as a company subscriber or to become a new subscriber. It should be emphasized one more time: the commitment function depends on the purpose of research. Different commitment function can provide totally different results and conclusions. In case of the number of calls and duration of calls difference in the social position ranking is not significant. Kendall's coefficient for the same $\varepsilon$ is very high $K(X,Y)=0.871$. Also the distribution of social position is almost the same for both commitment function (section 6.2 Distribution of Social Position). On the other hand, the research has revealed that there are 10% less duplicates when the commitment function, used in the process of social position assessment, was evaluated based on the duration of phone calls than based on the number of phone calls (section 6.5.2 Distribution and Number of Duplicates) and it means that commitment function based on duration of phone calls is better to diversify users. These observations show the difference between various commitment functions and answer the research question no. 3.

To answer research question no.5 and find key users, by utilizing social position measure, three different SPIN algorithms were proposed $SPIN^{nodes}$, $SPIN^{edges}$ and $SPIN^{hybrid}$. Each of those algorithms presents different way how to calculate social position (research question no. 4). The main difference between SPIN algorithms is the approach to calculation from the "node perspective", the "edge perspective" and the "mixed perspective" which join both node and edge perspective. For detailed information about the difference between SPIN algorithm variants see section 3.3 The SPIN Algorithm. Of course, all three algorithms provide the same results. The difference between various methods of social position calculation is their processing time. The comparison between SPIN$^{nodes}$, SPIN$^{edges}$ and SPIN$^{hybrid}$ shows that if the computational time is taken into consideration the SPIN$^{edges}$ is unquestionably the best algorithm. It is at least 120 times faster than SPIN$^{nodes}$ algorithm and at least 2 times faster than the SPIN$^{hybrid}$. For complete algorithms comparison and answer for research question no. 6 see section 6.4 Efficiency Tests.

The conducted research indicates that answer for research question no. 5 is yes, social position can be utilized to extract key users. SPIN$^{edges}$ was used to calculate user social position for the following $\varepsilon$ values: 0.1, 0.2, 0.3, 0.4, 0.5, 0.6, 0.7, 0.8, 0.9. For each $\varepsilon$ the ranking of first 15 users is almost the same regardless of commitment function (see section 6.1 The Calculation of Social Position). Among these first 15 users there are only big institutions: 4 hospitals, 3 banks or financial institution, 2 political-administrative institution, one industrial company, one the protection of dogs foundation, one affiliation of train hauliers and one car hire company in the centre of London[4].

In order to answer for research question no. 7 social position measure was compared with two others centrality measures indegree IC or outdegree OC (see section 6.5 Social Position versus Other Centrality Indices). Processing time of the fastest version of social position algorithm is approximately the same as an IC or OC centrality calculation time and they are the simplest to calculate measures from centrality indicates. Note that closeness and betweenness are much more difficult to compute e.g. shortest path problem in directed and weighted network for closeness and betweenness. Next fact that should be emphasized is that IC and OC take into consideration only the first level neighbours whereas the social position of user depends on the positions of all users within the network. This causes that the social position measure is much more diverse than other centrality indices. Moreover IC and OC generates much more duplicates than social position. In the analysed network there are 4,455,802 nodes and it means that around 2 mln users have distinct values of social positions whereas only 1,640 users obtained different indegree centrality and 1,662 different outdegree centrality. For that reason, the percentage of duplicates exceeds 99% for degree measures whereas it is around 60% for social position measure where commitment was evaluated based on the number of phone calls and 50% for social position measure where commitment was evaluated based on the duration of phone calls (section 6.5.2 Distribution and Number of Duplicates). So social position measure is very good in diversifying users, define user position based on positions of all users within the network and doing this during the same time as the simplest centrality indicates. Additionally, social position ranking can change depending on commitment function when IC and OC rankings are always the same so many different rankings can be made based on one network. Thanks to all this advantages social position appears to be powerful and one of the best measure which can be used to extract key users not only from telecommunication data.

Further work will focus on comparing the social position with all others centrality indicates to prove that the social position not only can be utilize to find key users but also is the effective measure to do that.

## Acknowledgement

This work was supported by MSc. Katarzyna Musiał, Ph.D. Przemysław Kazienko – Wrocław University of Technology, Institute of Applied Informatics and Social Network Group @ Wrocław University of Technology.

---

[4] Due to security and users protection issues more detailed information about users and their exact names and position in ranking cannot be reviled.

# 8 APPENDIX

## 8.1 Application which use social network analysis

1. Uncloak criminal network amongst slumlords
2. Reveal how hospital-acquired infection[HAI] spread amongst patients and medical staff
3. Map and weave nationwide volunteer network
4. Improve the innovation of a group of scientists and researchers in a worldwide organization
5. Find emergent leaders in fast growing company
6. Improve leadership and team chemistry for sports franchises
7. Build a grass roots political campaign
8. Determine influential journalists and analysts in the IT industry
9. Unmask the spread of HIV in a prison system
10. Map executive's personal network based on email flows
11. Discover the network of Innovators in a regional economy
12. Analyze book selling patterns to position a new book
13. Map a group of entrepreneurs in a specific marketspace
14. Find an organization's go-to people in various knowledge domains
15. Map interactions amongst blogs on various topics
16. Reveal key players in an investigative news story
17. Map national network of professionals involved in a change effort
18. Improve the functioning of various project teams
19. Map communities of expertise in various medical fields
20. Help large organization locate employees in new buildings
21. Examine a network of farm animals to analyze how disease spreads from one cow to another
22. Map network of Jazz musicians based on musical styles and CD sales
23. Discover emergent communities of interest amongst faculty at various universities
24. Reveal cross-border knowledge flows based on research publications
25. Expose business ties & financial flows to investigate possible criminal behavior
26. Uncover network of characters in a fictional work
27. Analyze managers' networks for succession planning
28. Discern useful patterns in clickstreams on the WWW
29. Locate technical experts and the paths to access them in engineering organization

Source: [27]

## 8.2 Tables Index

## 8.3 Figures Index